\documentstyle[preprint,eqsecnum,aps,epsf,epsfig]{revtex}
\tightenlines

\begin{document}
\draft

\title{Critical exponents for diluted resistor networks
}
\author{O. Stenull, H. K. Janssen and K. Oerding
}
\address{
Institut f\"{u}r Theoretische Physik III\\Heinrich--Heine--Universit\"{a}t\\Universit\"{a}tsstra{\ss}e 1\\40225 D\"{u}sseldorf, Germany
}
\date{\today}
\maketitle

\begin{abstract}
An approach by Stephen is used to investigate the critical properties of randomly diluted resistor networks near the percolation threshold by means of renormalized field theory. We reformulate an existing field theory by Harris and Lubensky. By a decomposition of the principal Feynman diagrams we obtain a type of diagrams which again can be interpreted as resistor networks. This new interpretation provides for an alternative way of evaluating the Feynman diagrams for random resistor networks. We calculate the resistance crossover exponent $\phi$ up to second order in $\epsilon=6-d$, where $d$ is the spatial dimension. Our result $\phi=1+\epsilon /42 +4\epsilon^2 /3087$ verifies a previous calculation by Lubensky and Wang, which itself was based on the Potts--model formulation of the random resistor network.
\end{abstract}

\narrowtext

\section{Introduction}
\label{introduction}

Percolation\cite{bunde_havlin_91_stauffer_aharony_92} is a leading paradigm for disorder. With regard to possible applications transport properties, e.g.\ electric conductance, are of particular interest in percolation theory. Electric transport on percolation clusters is commonly modelled by a random resistor network. In this model, bonds on a $d$-dimensional lattice are occupied with probability $p$ and unoccupied with probability $1-p$. Each occupied bond has a finite nonzero conductance $\sigma$ whereas unoccupied bonds have conductance zero. Suppose the system is near the percolation threshold, i.e.\ $p$ is close to the critical concentration $p_c$ above which an infinite cluster exists. If one measures the resistance $R( {\rm{\bf x}} ,{\rm{\bf x}}^\prime )$ between two lattice sites ${\rm{\bf x}}$ and ${\rm{\bf x}}^\prime$ known to be on the same cluster, one finds that the average over all possible configurations $M^1_R$ obeys\cite{harris_fisch_77,dasgupta_harris_lubensky_78} $M^1_R \sim |{\rm{\bf x}} - {\rm{\bf x}}^\prime |^{\phi / \nu}$, where $\nu $ is the correlation length exponent defined by $\xi \sim (p-p_c)^{-\nu}$.
The entire probability distribution for the resistance scales with the single exponent $\phi$\cite{rammal_lemieux_tremlay_85,harris_lubensky_85}.

The theory of random resistor networks broke ground in the 70's\cite{last_thouless_71,skal_shklovskii_74,straley_76,ephross_shklovskii_76,degennes_76,straley_77,alexander_orbach_82}. Kasteleyn and Fortuin\cite{fortuin_kasteleyn_72} realized a connection between the random resistor network and the $q\to 0$ limit of the $q$--state Potts--model. Stephen\cite{stephen_78} found an approach connected to the $xy$--model. A  Potts--model based formulation\cite{harris_fisch_77,harris_lubensky_87a} enabled Lubensky and Wang\cite{lubensky_wang_85} to calculate the crossover exponent $\phi$ up to second order in $\epsilon =6-d$, $d$ being the spatial dimension, as $\phi=1+\epsilon /42 +4\epsilon^2 /3087$. Nevertheless the approach by Stephen has been more fruitful\cite{harris_lubensky_87b}. For example, it has been employed to calculate $\phi$ for the random resistor network up to first order in $\epsilon$\cite{harris_lubensky_87b}, several crossover exponents for a diluted network of Josephson junctions\cite{john_lubensky_85}, $\phi$ for a network with a singular distribution of resistances\cite{lubensky_tremblay_86}, noise exponents for the random resistor network with fluctuating conductances\cite{park_harris_lubensky_87} as well as crossover and noise exponents for a randomly diluted network of nonlinear resistors\cite{harris_87}. However, we are not aware of any work up to now calculating $\phi$ based on this approach up to second order in $\epsilon$.

The state of the art in the theory of random resistor networks dates back to the 80's. Since then progress has been reported only rarely, e.g.\ by Fourcade and Tremblay\cite{fourcade_tremblay_95}. The complexity that the field theory of random resistor networks has reached might be a reason for this. Here we reformulate a field theory by Harris and Lubensky\cite{harris_lubensky_87b} based on Stephen in a way, that we believe is less complex and more intuitive. We present in detail a calculation of $\phi$ up to second order in $\epsilon$. We hope this fosters further calculations of this type which might appear in the future.

\section{The Model}
\label{model}
In this section we line out the essentials of the derivation of a field theory for random resistor networks due to Stephen\cite{stephen_78} and Harris and Lubensky\cite{harris_lubensky_87b}. We will provide the reader with indispensable background and clarify certain points.

\subsection{Kirchhoff's equations}
\label{kirchhoffsEquations}
Consider a $d$--dimensional lattice. Each bond is occupied by a resistor of conductance $\sigma$ with probability $p$ or unoccupied with probability $1-p$. Moreover each lattice site is connected to ground by a capacitor. For convenience all capacitors are taken to have the value unity. Kirchhoff's law applied to site $i$ reads
\begin{eqnarray}
\label{kirchhoff}
\dot{Q_i}=\sum_{j}{ \sigma_{i ,j} \left( V_{j} - V_i \right) } + I_i = i \omega V_i \ ,
\end{eqnarray}
where $Q_i$ and $V_i$ are the charge and the potential at site $i$ and $\omega$ is the corresponding frequency. $I_i$ is an externally imposed current and the sum runs over all nearest neighbors. In order to obtain a solution for the voltages Eq.~(\ref{kirchhoff}) may be cast into matrix form,
\begin{eqnarray}
\underline{\underline{S}} \ \underline{V} = \underline{I} \ ,
\end{eqnarray}
by setting $S_{i \neq j} = -\sigma_{i ,j}$ and $S_{i ,i}=i\omega + \sum_{j}{\sigma_{i ,j}}$. For $\omega \neq 0$ the inverse of $\underline{\underline{S}}$ is well defined and we can write
\begin{eqnarray}
\underline{V} = \underline{\underline{S}}^{-1} \underline{I} \ .
\end{eqnarray}
The limit $\omega \to 0 $ requires some cautiousness. To characterize clusters
${\mathcal C \mathnormal}$ we define vectors $\underline{e}({\mathcal C \mathnormal} )$ with $e_{i} ({\mathcal C \mathnormal} ) = \mbox{0} $ if $i \notin {\mathcal C \mathnormal}$ and $e_i ({\mathcal C \mathnormal} ) = N({\mathcal C \mathnormal} )^{-1/2}$ if $i \in {\mathcal C \mathnormal}$, where $N({\mathcal C \mathnormal} )$ denotes the number of sites belonging to ${\mathcal C \mathnormal}$. The vectors $\underline{e}$ are eigenvectors of $\underline{\underline{S}}$:
\begin{eqnarray}
\underline{\underline{S}} \ \underline{e}({\mathcal C \mathnormal} ) = i \omega \underline{e}({\mathcal C \mathnormal} ) \ .
\end{eqnarray}
Thus the inverse of $\underline{\underline{S}}$ may be written as
\begin{eqnarray}
\label{inverseOfD}
S^{-1}_{i ,j} = \frac{z_{i ,j}}{i \omega} + \tilde{S}^{-1}_{i ,j} \ ,
\end{eqnarray}
with $\tilde{\underline{\underline{S}}}^{-1}$ summarizing the part of $\underline{\underline{S}}^{-1}$ that stays finite in the limit $\omega \to 0$, $z_{i ,j}=N({\mathcal C \mathnormal} )^{-1}$ if $i ,j \in {\mathcal C \mathnormal}$ and $z_{i ,j}=0$ if $i$ and $j$ are not located on the same cluster. From Eq.~(\ref{inverseOfD}) we conclude that singularities due to zero modes can be regularized as far as the infinite cluster is concerned by taking the thermodynamic limit before $\omega \to 0$. In the remainder of this article we focus on vanishing $\omega$ unless stated otherwise.

Suppose a current $I$ is put into a cluster at site $x$ and taken out at site $x^\prime$. Those sites connected to both $x$ and $x^\prime$ by two mutually non intersecting paths are constituting the backbone between $x$ and $x^\prime$. The current at a site $i$ belonging to the backbone may be written as
\begin{eqnarray}
I_i = I \left( \delta_{i,x} - \delta_{i,x^\prime} \right)
\end{eqnarray}
and the voltage at site $i$ is simply
\begin{eqnarray}
V_{i} = \left( S_{i ,x}^{-1} - S_{i,x^\prime}^{-1} \right) I \ .
\end{eqnarray}
The difference in voltage between the source and the sink is
\begin{eqnarray}
V_{x} - V_{x^\prime} = \left[ S_{x ,x}^{-1} + S_{x^\prime ,x^\prime}^{-1} - 2S_{x ,x^\prime}^{-1} \right] I \ ,
\end{eqnarray}
which states that the resistance $R(x ,x^\prime)$ between sites $x$ and $x^\prime$ reads
\begin{eqnarray}
R(x ,x^\prime) = S_{x ,x}^{-1} + S_{x^\prime ,x^\prime}^{-1} - 2S_{x ,x^\prime}^{-1} \ .
\end{eqnarray}
Different cases may be distinguished: If $x$ and $x^\prime$ lay on the same cluster then
\begin{eqnarray}
R(x ,x^\prime) = \tilde{S}_{x ,x}^{-1} + \tilde{S}_{x^\prime ,x^\prime}^{-1} - 2\tilde{S}_{x ,x^\prime}^{-1} \ .
\end{eqnarray}
If $x$ and $x^\prime$ are not connected then $\tilde{S}_{x ,x^\prime}^{-1}=0$ and hence
\begin{eqnarray}
R(x ,x^\prime) = \frac{1}{i\omega} \left( z_{x ,x} + z_{x^\prime ,x^\prime} \right) + \tilde{S}_{x ,x}^{-1} + \tilde{S}_{x^\prime ,x^\prime}^{-1} \ .
\end{eqnarray}
We see that $R(x ,x^\prime) \to \infty$ for $\omega \to 0$ if $x$ and $x^\prime$ are not on the same cluster. In the limit of widely separated sites one has $S^{-1}_{x ,x^\prime} \to 0$. Thus the resistance $R_\infty$ between two infinitely separated points on the infinite cluster is
\begin{eqnarray}
R_\infty = \tilde{S}_{x ,x}^{-1} + \tilde{S}_{x^\prime ,x^\prime}^{-1} = 2 \tilde{S}_{x ,x}^{-1} \ ,
\end{eqnarray}
with the last equality holding for homogenous systems.

Now we turn to the power $P=I \left( V_x - V_{x^\prime} \right)$ dissipated on the backbone. It may be written as
\begin{eqnarray}
\label{power1}
P = R(x ,x^\prime)^{-1} \left( V_x - V_{x^\prime} \right)^2 = \sum_{k,l} \sigma_{k,l} \left( V_k - V_l \right)^2 = P \left( \left\{ V \right\} \right)
\end{eqnarray}
with the sum running over all nearest neighbor pairs on the backbone and $\left\{ V \right\}$ denoting the corresponding set of voltages. In terms of $P$ Kirchhoff's equation (\ref{kirchhoff}) arises as a consequence of the variation principle
\begin{eqnarray}
\frac{\partial}{\partial V_i} \left[ \frac{1}{2} P \left( \left\{ V \right\} \right) + \sum_k I_k V_k \right] = 0 \ .
\end{eqnarray}

Equivalent to Eq.~(\ref{power1}) the power can be expressed in terms of the currents as
\begin{eqnarray}
\label{power2}
P = R\left( x ,x^\prime \right) I^2 = \sum_{k,l} \sigma^{-1}_{k,l} I_{k,l}^2 \ .
\end{eqnarray}
Suppose the backbone contains closed loops as sub--networks with currents $\left\{ I^{(l)} \right\}$ circulating independently around these loops. Then the current flowing through a certain bond is not only a function of $I$ but also of the set of loop currents:
\begin{eqnarray}
I_{k,l} = I_{k,l} \left( \left\{ I^{(l)} \right\} , I \right) \ .
\end{eqnarray}
Conservation of charge holds for every ramification of the backbone and this gives rise to another variation principle:
\begin{eqnarray}
\label{variationPrinciple2}
\frac{\partial}{\partial I^{(l)}} P \left( \left\{ I^{(l)} \right\} , I \right) = 0 \ .
\end{eqnarray} 
Eq.~(\ref{variationPrinciple2}) may be used to eliminate the loop currents and thus provides us with a method to determine the total resistance of the backbone via Eq.~(\ref{power2}).
 
\subsection{Replica formalism}
\label{replicaFormalism}
From the discussion above it is evident that our task is in principle to invert a random matrix. This inversion can be generated by gaussian integration. However, we are interested in the average resistance $\langle R(x ,x^\prime) \rangle_{C}$ and hence the average over all possible realizations of the diluted configuration for fixed $p$ remains to be performed. This can be achieved by employing replica technique\cite{mezard_parizi_virasoro_87}. The network is replicated $D$--fold: $\underline{V} \to \vec{\underline{V}} = \left( \underline{V}^{(1)}, \ldots , \underline{V}^{(D)} \right)$. The replication procedure leeds to the extended generating function
\begin{eqnarray}
\label{genFkt}
\left\langle Z^{-D} \int \prod_i \prod_{\alpha =1}^D dV_i^\alpha \exp \left( -\frac{1}{2} P + i \vec{\underline{\lambda}} \cdot \vec{\underline{V}} \right) \right\rangle_C = \left\langle \exp \left( -\frac{1}{2} \vec{\underline{\lambda}} \cdot \underline{\underline{S}}^{-1} \vec{\underline{\lambda}} \right) \right\rangle_C \ ,
\end{eqnarray}
with the power
\begin{eqnarray}
\label{power}
P = \vec{\underline{V}} \cdot \underline{\underline{S}} \  \vec{\underline{V}} = \sum_{i,j,\alpha } V_i^{(\alpha )} S_{i,j} V_j^{(\alpha )}
\end{eqnarray}
and where $\vec{\underline{\lambda}} \cdot \vec{\underline{V}} = \sum_{i, \alpha} \lambda_i^{(\alpha )} V_i^{(\alpha )}$. The normalization constant $Z$ is adjusted in the usual way:
\begin{eqnarray}
\label{norm}
Z = \int \prod_{i} dV_{i} \exp \left( -\frac{1}{2} \underline{V} \cdot \underline{\underline{S}} \ \underline{V} \right) \ .
\end{eqnarray} 
The choice $\vec{\lambda}_i = \vec{\lambda} \left( \delta_{i,x} - \delta_{i,x^\prime} \right)$ provides us with a generating function for the average resistance:
\begin{eqnarray}
\label{repGenFkt}
\left\langle \psi_{\vec{\lambda}}(x)\psi_{-\vec{\lambda}}(x^\prime) \right\rangle_{\mbox{\scriptsize{rep}}} = \left\langle \exp \left( -\frac{\vec{\lambda}^2}{2} R(x ,x^\prime ) \right) \right\rangle_C \ ,
\end{eqnarray}
where
\begin{eqnarray}
\psi_{\vec{\lambda}}(x) = \exp \left( i \vec{\lambda} \cdot \vec{V}_x \right)  \ , \vec{\lambda} \neq \vec{0}
\end{eqnarray}
and $\langle ... \rangle_{\mbox{\scriptsize{rep}}}$ denotes the average over all replicated voltage and lattice configurations. An expression similar to Eq.~(\ref{repGenFkt}) holds for the resistance to infinity:
\begin{eqnarray}
\left\langle \psi_{\vec{\lambda}}(x) \right\rangle_{\mbox{\scriptsize{rep}}} = \left\langle \exp \left( -\frac{\vec{\lambda}^2}{2} R_\infty \right) \right\rangle_C \ .
\end{eqnarray}

As we shall see later on, it is useful to define an indicator function that is unity if $x$ and $x^\prime$ lay on the same cluster and zero otherwise. Consider the limit $\sigma \to \infty$. From the discussion in subsection \ref{kirchhoffsEquations} it is clear that $R \left( x, x^\prime \right) = 0$ if $x$ and $x^\prime$ are connected and $R \left( x, x^\prime \right) = \infty$ else. This suggests to define the indicator function in terms of the average $\langle \psi_{\vec{\lambda}}(x)\psi_{-\vec{\lambda}}(x^\prime) \rangle$ over all replicated voltage configurations for a given realization of the diluted lattice as\begin{eqnarray}
\chi (x ,x^\prime) = \lim_{\sigma \to \infty} \langle \psi_{\vec{\lambda}} (x) \psi_{-\vec{\lambda}} (x^\prime) \rangle \ .
\end{eqnarray}
Similarly
\begin{eqnarray}
\chi (x ) = \lim_{\sigma \to \infty} \langle \psi_{\vec{\lambda}} (x) \rangle 
\end{eqnarray}
indicates if $x$ is located on the infinite cluster.

Some remarks should be made at this point. Since infinite voltage drops between different clusters may occur it is not guaranteed that $Z$ stays finite, i.e.\ the limit $\lim_{D \to 0} Z^D$ is not well defined. This problem can be regularized by switching to voltage variables $\vec{\theta}$ taking discrete values  on a $D$--dimensional torus. The voltages are discretized by setting $\vec{\theta} = \Delta \theta \vec{k}$, where $\Delta \theta = \theta_M /M$ is the gap between successive voltages, $\theta_M$ is a voltage cutoff, $\vec{k}$ is a $D$--dimensional and $M$ a positive integer. The components of $\vec{k}$ are restricted to $-M < k^{(\alpha)} \leq M$ and periodic boundary conditions are generated by equating $k^{(\alpha )}=k^{(\alpha )} \mbox{mod} (2M)$. The continuum may be restored by taking $\theta_M \to \infty$ and $\Delta \theta \to 0$. By setting $\theta_M = \theta_0 M$, $M=m^2$ and respectively $\Delta \theta = \theta_0 /m$ the two limits can be taken simultaneously via $m \to \infty$. Note that the limit $D \to 0$ has to be taken before any other limit. Since the voltages and $\vec{\lambda}$ are conjugated variables, $\vec{\lambda}$ is affected by the discretization as well:
\begin{eqnarray}
\vec{\lambda} = \Delta \lambda \vec{l} \ , \ \Delta \lambda \Delta \theta = \pi \ ,
\end{eqnarray}
where $\vec{l}$ is a $D$--dimensional integer taking the same values as $\vec{k}$. This choice guarantees that the completeness and orthogonality relations
\begin{mathletters}
\label{complete}
\begin{eqnarray}
\frac{1}{(2M)^D} \sum_{\vec{\theta}} \exp \left( i \vec{\lambda} \cdot \vec{\theta} \right) = \delta_{\vec{\lambda} ,\vec{0} \hspace{0.15em}\mbox{\scriptsize{mod}}(2M \Delta \lambda) }
\end{eqnarray}
and
\begin{eqnarray}
\frac{1}{(2M)^D} \sum_{\vec{\lambda}} \exp \left( i \vec{\lambda} \cdot \vec{\theta} \right) = \delta_{\vec{\theta} ,\vec{0} \hspace{0.15em}\mbox{\scriptsize{mod}}(2M \Delta \theta)}
\end{eqnarray}
\end{mathletters}
do hold. Eq.~(\ref{complete}) provides us with the Fourier transform \begin{eqnarray}
\label{fouriertransform}
\Phi_{\vec{\theta}} \left( x \right) = (2M)^{-D} \sum_{\vec{\lambda} \neq \vec{0}} \exp \left( i \vec{\lambda} \cdot \vec{\theta} \right) \psi_{\vec{\lambda}} (x) = \delta_{\vec{\theta}, \vec{\theta}_{x}} - (2M)^{-D} \end{eqnarray}
with the condition
\begin{eqnarray}
\sum_{\vec{\theta }} \Phi_{\vec{\theta}} \left( x \right) = 0 \ .
\end{eqnarray}
Note that $\Phi$ is nothing else than a Potts--spin\cite{Zia_Wallace_75} with $q=(2M)^D$ states.

\subsection{Field theoretic Hamiltonian}
\label{fieldTheoreticHamiltonian}
We proceed with the evaluation of Eq.~(\ref{genFkt}). Carrying out the average over the diluted lattice configurations provides us with the weight $\exp (-H_{\mbox{\scriptsize{rep}}})$ of the average $\langle ... \rangle_{\mbox{\scriptsize{rep}}}$:
\begin{eqnarray}
H_{\mbox{\scriptsize{rep}}} = - \ln \left\langle  \exp \left( - \frac{1}{2} P \right) \right\rangle_C = \frac{i \omega}{2} \sum_{i} \vec{\theta}_{i} \cdot \vec{\theta}_{i} - \sum_{i ,j} \ln \left\langle  \exp \left( - \frac{1}{2} \sigma_{i ,j} \left( \vec{\theta}_{i} - \vec{\theta}_{j} \right) \cdot \left( \vec{\theta}_{i} - \vec{\theta}_{j} \right) \right)  \right\rangle_C \ .
\end{eqnarray}
By dropping a constant term $N_B \ln (1-p)$ with $N_B$ being the number of bonds in the undiluted lattice we obtain
\begin{eqnarray}
H_{\mbox{\scriptsize{rep}}} = \sum_{i} h \left( \vec{\theta}_{i} \right) + \sum_{i ,j} K \left( \vec{\theta}_{i} - \vec{\theta}_{j} \right) \ ,
\end{eqnarray}
where
\begin{eqnarray}
\label{breakingField}
h \left( \vec{\theta} \right) = \frac{i \omega}{2} \vec{\theta} \cdot \vec{\theta} 
\end{eqnarray}
and
\begin{eqnarray}
K \left( \vec{\theta} \right) = - \ln \left\{ 1 + \frac{p}{1-p} \exp \left( - \frac{1}{2} \sigma \vec{\theta} \cdot \vec{\theta} \right) \right\} \ .
\end{eqnarray}
The Fourier transform of $K$,
\begin{eqnarray}
\tilde{K} \left( \vec{\lambda} \right) = - \frac{1}{(2M)^D} \sum_{\vec{\theta}} \exp \left( -i \vec{\lambda} \cdot \vec{\theta} \right) \ln \left( 1 + \frac{p}{1-p} \exp \left( - \frac{1}{2} \sigma \vec{\theta} \cdot \vec{\theta} \right) \right)
\end{eqnarray}
can be Taylor expanded as 
\begin{eqnarray}
\label{taylorExp}
\tilde{K} \left( \vec{\lambda} \right) = \tau + \sum_{p=1}^{\infty} w_p \left( \vec{\lambda}^2 \right)^p \ ,
\end{eqnarray}
with $\tau$ and $w_p \sim \sigma^{-p}$ being expansion coefficients. Since $K$ decays exponentially in $\sigma \vec{\theta}^2$ the series (\ref{taylorExp}) may be terminated after the quadratic term for large $\sigma$.  Further on we omit factors $(2M)^{-D}$ which become unity in the limit $D \to 0$. We define the discrete nabla operator $\nabla_{\vec{\theta}}$ through
\begin{eqnarray}
- \sum_{\vec{\theta}} \nabla_{\vec{\theta}} \Phi_{\vec{\theta}} \left( x \right) \cdot \nabla_{\vec{\theta}} \Phi_{\vec{\theta}} \left( x \right) = \sum_{\vec{\lambda} \neq \vec{0}} \vec{\lambda}^2 \psi_{\vec{\lambda}}(x) \psi_{-\vec{\lambda}}(x)
\end{eqnarray}
and obtain
\begin{eqnarray}
\label{Kequals}
 K \left( \Delta_{\vec{\theta}} \right) = \tau - w \Delta_{\vec{\theta}} \ ,
\end{eqnarray}
where $w = w_1$. Similarly the right hand side of Eq.~(\ref{breakingField}) may be viewed as the leading term of an expansion
\begin{eqnarray}
\label{hequals}
h \left( \vec{\theta} \right) = \sum_{p=1}^\infty h_p \left( \vec{\theta}^2 \right)^p 
\end{eqnarray}
with $h_1 = i \omega /2$.

Now we can set up a field theoretic Hamiltonian $\mathcal H \mathnormal$ in compliance with the symmetries of the model. The average over the configurations of the diluted lattice renders the model symmetric under spatial translations and rotations. Another feature of the model is that it is local since only nearest neighbors enter in Eq.~(\ref{kirchhoff}). In the limit of perfect transport ($\sigma \to \infty $) and in the absence of external fields ($h_p =0$) the model is invariant against permutations of all $q=(2M)^D$ states of the Potts--spins $\Phi_{\vec{\theta}} \left( x \right)$. If one allows $w_p \neq 0$ this $S_q$ symmetry is lost. The model remains gauge invariant under a shift of the voltages by an arbitrary finite potential as can be inferred from Eq.~(\ref{kirchhoff}). This symmetry corresponds to translational invariance in replica space: $\Phi_{\vec{\theta}} \left( x \right) \leftrightarrow \Phi_{\vec{\theta} - \vec{\theta}_0} \left( x \right)$. Moreover we deduce from the quadratic form of the power in Eq.~(\ref{power}) that the model possesses of an $O(D)$ symmetry in replica space. Additionally admitting $h_p \neq 0$ results in breaking of the gauge invariance. 

We proceed with the usual coarse graining step and replace the Potts--spins $\Phi_{\vec{\theta}} \left( x \right)$ by the order parameter field $\varphi \left( {\rm{\bf x}} ,\vec{\theta} \right)$. By constructing all possible invariants of the symmetries discussed above from $\sum_{\vec{\theta}} \varphi \left( {\rm{\bf x}} ,\vec{\theta} \right)^p$ and gradients thereof the following Hamiltonian in spirit of the Landau--Ginzburg--Wilson functional is obtained:
\begin{eqnarray}
\label{fieldhamiltonian}
\mathcal H \mathnormal = \int d^dx \sum_{\vec{\theta}} \left\{ \frac{1}{2} \varphi \left( {\rm{\bf x}} , \vec{\theta} \right) K \left( \Delta ,\Delta_{\vec{\theta}} \right) \varphi \left( {\rm{\bf x}} , \vec{\theta} \right) + \frac{g^3}{6}\varphi \left( {\rm{\bf x}} , \vec{\theta} \right)^3 + h \left( {\rm{\bf x}} ,\vec{\theta} \right) \varphi \left( {\rm{\bf x}} , \vec{\theta} \right)
\right\} \ , 
\end{eqnarray}
where terms of higher order in the fields have been neglected since they turn out to be irrelevant. The coarse grained kernel and external field must resemble the features of the original $K$ and $h$. This gives
\begin{eqnarray}
\label{finalHamil}
\mathcal H \mathnormal &=& \int d^dx \sum_{\vec{\theta}} \left\{ \frac{\tau}{2} \varphi \left( {\rm{\bf x}} , \vec{\theta} \right)^2 + \frac{w}{2} \left( \nabla_{\vec{\theta}} \varphi \left( {\rm{\bf x}} , \vec{\theta} \right) \right)^2 + \frac{1}{2} \left( \nabla \varphi \left( {\rm{\bf x}} , \vec{\theta} \right) \right)^2 \right.
\nonumber \\
&+& \left. \frac{g^3}{6}\varphi \left( {\rm{\bf x}} , \vec{\theta} \right)^3 + \frac{i \omega}{2} \vec{\theta}^2 \varphi \left( {\rm{\bf x}} , \vec{\theta} \right)
\right\} \ ,
\end{eqnarray}
where $\tau$, $w$ and $\omega$ are now coarse grained analogues of the original coefficients. Note that $\mathcal H \mathnormal$ reduces to the usual $(2M)^D$--states Potts--model Hamiltonian by setting $w=h=0$ as one one retrieves purely geometrical percolation in the limit of vanishing $h$ and $w$ ($\sigma \to \infty$). 

It is worth pointing out that the problem of calculating the moments of the resistance distribution has been reshuffled: $M^k_R = \left\langle \chi({\rm{\bf x}}, {\rm{\bf x}}^\prime) R^k({\rm{\bf x}} ,{\rm{\bf x}}^\prime )\right\rangle_C$ may be obtained by taking the $k$--th derivative of
\begin{eqnarray}
\label{reshuffled}
\lim_{D \to 0} \left\langle \psi_{\vec{\lambda}}({\rm{\bf x}}) \psi_{-\vec{\lambda}}({\rm{\bf x}}) \right\rangle_{\mathcal H \mathnormal} &=& \left\langle \chi({\rm{\bf x}}, {\rm{\bf x}}^\prime) \right\rangle_C - \frac{\vec{\lambda}^2}{2}\left\langle \chi({\rm{\bf x}}, {\rm{\bf x}}^\prime) R({\rm{\bf x}}, {\rm{\bf x}}^\prime) \right\rangle_C + \ldots
\nonumber \\ 
&\ldots& + \frac{1}{k!} \left( - \frac{\vec{\lambda}^2}{2} \right)^k \left\langle \chi({\rm{\bf x}}, {\rm{\bf x}}^\prime) R^k({\rm{\bf x}}, {\rm{\bf x}}^\prime) \right\rangle_C + \ldots
\end{eqnarray}
with respect to $\vec{\lambda}^2$ at $\vec{\lambda}^2 = 0$.

\subsection{Scaling properties}
\label{scalingProperties}
We conclude this section with a scaling analyses of the Hamiltonian~(\ref{fieldhamiltonian}). Let $P$ denote the set of parameters $\left\{ w_p , h_p \right\}$ and $b$ some scaling factor for the voltage variable: $\vec{\theta} \to b \vec{\theta}$. By substitution of $\varphi \left( {\rm{\bf x}} , \vec{\theta} \right) = \varphi^\prime \left( {\rm{\bf x}} , b\vec{\theta} \right)$ into the Hamiltonian we get
\begin{eqnarray}
\label{scaling1}
\mathcal H \mathnormal \left[ \varphi^\prime \left( {\rm{\bf x}} , b \vec{\theta} \right) , P \right] &=& \int d^dx \sum_{\vec{\theta}} \left\{ \frac{1}{2} \varphi^\prime \left( {\rm{\bf x}} , b \vec{\theta} \right) K \left( \Delta ,\Delta_{\vec{\theta}} \right) \varphi^\prime \left( {\rm{\bf x}} , b \vec{\theta} \right) + \frac{g^3}{6}\varphi^{\prime } \left( {\rm{\bf x}} , b \vec{\theta} \right)^3 \right.
\nonumber \\
&+& h \left(\vec{\theta} \right) \varphi^\prime \left( {\rm{\bf x}} , b \vec{\theta} \right)
\Bigg\} \ .
\end{eqnarray}
Renaming the scaled voltage variables $\vec{\theta}^\prime = b \vec{\theta}$ leads to
\begin{eqnarray}
\label{scaling2}
\mathcal H \mathnormal \left[ \varphi^\prime \left( {\rm{\bf x}} , \vec{\theta}^\prime \right) , P \right] &=& \int d^dx \sum_{\vec{\theta}^\prime} \left\{ \frac{1}{2} \varphi^\prime \left( {\rm{\bf x}} , \vec{\theta}^\prime \right) K \left( \Delta , b^2 \Delta_{\vec{\theta}^\prime} \right) \varphi^\prime \left( {\rm{\bf x}} , \vec{\theta}^\prime \right) + \frac{g^3}{6}\varphi^{\prime } \left( {\rm{\bf x}} , \vec{\theta}^\prime \right)^3 \right.
\nonumber \\
&+& h \left( b^{-1} \vec{\theta}^\prime \right) \varphi^\prime \left( {\rm{\bf x}} , \vec{\theta}^\prime \right)
\Bigg\} \ .
\end{eqnarray}
Clearly a scaling of the voltage variable results in a scaling of the voltage cutoff: $\theta_M \to b \theta_M$. However, by taking the limit $D \to 0$ and then $m \to \infty$ the dependance of the theory on the cutoff drops out. No ultraviolet divergencies are encountered in integrations over voltage variables and hence the voltages are no origin of anomalous scaling. We can identify $\vec{\theta}^\prime$ and $\vec{\theta}$ and thus 
\begin{eqnarray}
\label{relForH}
\mathcal H \mathnormal \left[ \varphi \left( {\rm{\bf x}} , b \vec{\theta} \right) , P \right] = \mathcal H \mathnormal \left[ \varphi \left( {\rm{\bf x}} , \vec{\theta} \right) , P^\prime \right] \ ,
\end{eqnarray}
where $P^\prime = \left\{ b^{2p}w_p , b^{-2p} h_p \right\}$. We can conclude the following implication of Eq.~(\ref{relForH}) on correlation functions
\begin{eqnarray}
\label{correl}
G_N \left( \left\{ {\rm{\bf x}} ,\vec{\theta} \right\} ; \tau , \left\{ w_p , h_p \right\}  \right) = \int \mathcal D \mathnormal \varphi \ \varphi \left( {\rm{\bf x}}_1 , \vec{\theta}_1 \right) \ldots \varphi \left( {\rm{\bf x}}_N , \vec{\theta}_N \right)  \exp \left( - \mathcal H \mathnormal \left[ \varphi \left( {\rm{\bf x}} , \vec{\theta} \right) , P \right] \right) \ ,
\end{eqnarray}
where $\mathcal D \mathnormal \varphi$ indicates an integration over the set of variables $\left\{ \varphi \left( {\rm{\bf x}} , \vec{\theta} \right) \right\}$ for all ${\rm{\bf x}}$ and $\vec{\theta}$:
\begin{eqnarray}
G_N \left( \left\{ {\rm{\bf x}} ,\vec{\theta} \right\} ; \tau , \left\{ w_p , h_p \right\}  \right) = G_N \left( \left\{ {\rm{\bf x}} , b \vec{\theta} \right\} ; \tau , \left\{  b^{2p}w_p , b^{-2p} h_p \right\}  \right) \ .
\end{eqnarray}
The two point correlation function $G_2$ is the Fourier transform of $ \left\langle \psi_{\vec{\lambda}}({\rm{\bf x}}) \psi_{-\vec{\lambda}}({\rm{\bf x}}) \right\rangle_{\mathcal H \mathnormal}$. We deduce from Eq.~(\ref{reshuffled}) that
\begin{eqnarray}
\vec{\lambda}^2 M^1_R \left( \left( {\rm{\bf x}} , {\rm{\bf x}}^\prime \right) ;\tau , \left\{ w_p ,h_p \right\} \right) = \left( b^{-1} \vec{\lambda}  \right)^2 M^1_R \left( \left( {\rm{\bf x}} , {\rm{\bf x}}^\prime \right) ;\tau , \left\{ b^{2p}w_p , b^{-2p} h_p \right\} \right) \ .
\end{eqnarray}
The freedom to choose $b$ has not been exploited yet. With the choice $b^2 = w^{-1}$ the previous scaling relation turns into
\begin{eqnarray}
\label{scaling_w}
M^1_R \left( \left( {\rm{\bf x}} , {\rm{\bf x}}^\prime \right) ;\tau , \left\{ w_p ,h_p \right\} \right) = w M^1_R \left( \left( {\rm{\bf x}} , {\rm{\bf x}}^\prime \right) ;\tau , \left\{ \frac{w_p}{w^p} , w^p h_p \right\} \right) \ .
\end{eqnarray}
For $p>1$ coupling constants $w_p$ only appear as $w_p /w^p$. The associated exponent $p \phi -\phi_p$ is of order $p-1$ as mean field analysis shows. This indicates that terms of order $w_p \vec{\lambda}^{2p}$ for $p>1$ give rise to corrections to scaling. We keep only the leading terms $w \vec{\lambda}^2$ and $h_1 = i \omega /2$. By virtue of $ w \sim \sigma^{-1}$ the resulting scaling relation reads 
\begin{eqnarray}
\label{scalingFkt}
M^1_R \left( \left( {\rm{\bf x}} , {\rm{\bf x}}^\prime \right) ;\tau , \sigma, \omega \right) = 
\sigma^{-1} f \left( \left( {\rm{\bf x}} , {\rm{\bf x}}^\prime \right) ;\tau , \omega / \sigma \right) \ , 
\end{eqnarray}
where $f$ is a scaling function of $\omega / \sigma$.

\section{Two Loop Calculation And Renormalization}
\label{2loop}

\subsection{Diagrammatic expansion}
\label{diagrammaticExpansion}
To perform the renormalization program we start out with a dimensional analysis. A trivial consequence of the fact that the Hamiltonian (\ref{finalHamil}) must be dimensionless is that the involved quantities have naive dimensions ${\rm{\bf x}} \sim \mu^{-1}$, $w \vec{\lambda}^2 \sim \mu^2$, $\psi \sim \mu^{(d-2)/2}$, $\tau \sim \mu^2$ and $g \sim \mu^{(6-d)/2}$, where $\mu$ is a convenient inverse length scale. The positive dimension of the coupling constant $g$ shows that it is relevant for $d < d_c = 6$.

The principal elements contributing to the diagrammatic expansion are easily gathered from the Hamiltonian, namely the vertex $-g$ and the propagator
\begin{eqnarray}
\label{propagatorDecomp}
\frac{ 1 - \delta_{\vec{\lambda}, \vec{0}}}{{\rm{\bf p}}^2 + \tau + w\vec{\lambda}^2} = \frac{1}{{\rm{\bf p}}^2 + \tau + w\vec{\lambda}^2} - \frac{\delta_{\vec{\lambda}, \vec{0}}}{{\rm{\bf p}}^2 + \tau} \ ,
\end{eqnarray}
which is displayed in Fig.~\ref{propagator}.

The superficial degree of divergence $\delta$ in any one--particle--irreducible diagram composed of these elements is $\delta = dL-2P$, where $L$ denotes the number of loops and $P$ the number of propagators. The topologic relations $3V=E+2P$ and $L=P-V+1$, with $E$ being the number of external legs and $V$ being the number of vertices, lead for $d=d_c$ to $\delta=6-2E$. Therefore the only superficially divergent vertex functions are $\Gamma_2$ and $\Gamma_3$ (see Fig.~\ref{gammas}).

\subsection{Feynman diagrams as resistor networks}
\label{interpretation}
We learn from Eq.~(\ref{propagatorDecomp}) that the principal propagator (bold) decomposes into two propagators. One of them is carrying replica currents and we refer to it as conducting. The other one is not carrying $\vec{\lambda}$'s and we call it insulating. This decomposition of the bold propagator allows for a schematic decomposition of bold diagrams into sums of diagrams consisting of conducting and insulating propagators.

There is an important feature of the decomposition scheme that we want to point out at the instance of the diagram displayed in Fig.~\ref{feature}. The diagram reads
\begin{eqnarray}
\label{decomp_2}
\frac{g^4}{2} \sum_{\vec{\kappa}} \int_{{\rm{\bf k}} {\rm{\bf q}}} \frac{1}{\tau + {\rm{\bf k}}^2 + w \vec{\lambda}^2} \ \frac{1}{\tau + {\rm{\bf k}}^2} \ \frac{1}{\tau + {\rm{\bf k}}^2} \ \frac{1}{\tau + ({\rm{\bf k}} + {\rm{\bf q}})^2 + w \vec{\kappa}^2} \ \frac{1}{\tau + {\rm{\bf q}}^2 + w \vec{\kappa}^2} \ ,
\end{eqnarray}
where $\int_{{\rm{\bf k}} {\rm{\bf q}} }$ is an abbreviation for $(2\pi)^{-d}\int d^d k d^d q$. Schwinger parametrization leads to
\begin{eqnarray}
\label{decomp_3}
\mbox{(\ref{decomp_2})} &=& \frac{g^4}{2} \sum_{\vec{\kappa}} \int_{{\rm{\bf k}} {\rm{\bf q}}} \int_0^\infty \prod_{i=1}^5 ds_i \exp \left( -\tau \sum_{i=1}^5 s_i - \left( s_1 + s_2 + s_3 \right) {\rm{\bf k}}^2 - s_4 \left( {\rm{\bf k}} + {\rm{\bf q}} \right)^2 - s_5 {\rm{\bf q}}^2 \right)
\nonumber \\ 
&\times& \exp \left( - \left( s_4 + s_5 \right) w \vec{\kappa}^2 - s_1 w  \vec{\lambda}^2  
\right) \ .
\end{eqnarray}
The sum over $\vec{\kappa}$ factorizes
\begin{eqnarray}
\sum_{\vec{\kappa}} \exp \left( - \left( s_4 + s_5 \right) w \vec{\kappa}^2 \right) = \left( \sum_{\kappa} \exp \left( - \left( s_4 + s_5 \right) w \kappa^2 \right) \right)^D
\end{eqnarray}
and hence becomes unity in the limit $D \to 0$. What we encounter here in our example is a particular instance of the diagrammatic rule formulated in Fig.~\ref{rule}.

We apply the decomposition scheme to all one and two loop diagrams. The result is displayed in Figs.~\ref{decompRound0}--\ref{decompTri2_3_4}.

From the decomposition a new interpretation of the Feynman diagrams emerges. They may be viewed as resistor networks themselves with conducting propagators corresponding to conductors and insulating propagators to open bonds. The Schwinger parameters $s_i$ correspond to resistances $\sigma_i^{-1}$ and the replica variables $-i \vec{\lambda}_i$ to currents. As conservation of charge holds for every ramification in a resistor network, the $\vec{\lambda}$'s are conserved in each vertex. The $\vec{\lambda}$--dependent part of a diagram can be expressed in terms of its power $P$:
\begin{eqnarray}
\exp \left( -w \sum_i s_i \vec{\lambda}^2_i \right) = \exp \left( w P \left( \vec{\lambda} , \left\{ \vec{\kappa} \right\} \right) \right) 
\end{eqnarray}
with $\vec{\lambda}_i = \vec{\lambda}_i \left( \vec{\lambda} , \left\{ \vec{\kappa} \right\} \right)$, where (appart from a factor $-i$) $\vec{\lambda}$ is an external current and $\left\{ \vec{\kappa} \right\}$ denotes the set of independent loop currents. 

The new interpretation suggests an alternative way of computing the Feynman diagrams. To evaluate sums over independent loop currents
\begin{eqnarray}
\label{toEvaluate}
\sum_{\left\{ \vec{\kappa} \right\}} \exp \left( w P \left( \vec{\lambda} , \left\{ \vec{\kappa} \right\} \right) \right) 
\end{eqnarray}
one can employ the saddle point method that is exact in our case since the power is quadratic in the currents. Note that the saddle point equation is nothing else than the variation principle stated in Eq.~(\ref{variationPrinciple2}). Thus solving the saddle point equations is equivalent to determining the total resistance $R \left( \left\{ s_i \right\} \right)$ of a diagram and the saddle point evaluation of (\ref{toEvaluate}) yields
\begin{eqnarray}
\exp \left( -R \left(  \left\{ s_i \right\} \right) w \vec{\lambda}^2 \right) \ . 
\end{eqnarray}
After a completion of squares in the momenta the momentum integrations are straightforward. Thereafter all diagrams are of the form
\begin{eqnarray}
I \left( {\rm{\bf p}}^2 , \vec{\lambda}^2 \right) &=& I_P \left( {\rm{\bf p}}^2 \right) - I_W \left( {\rm{\bf p}}^2 \right) w \vec{\lambda}^2 + \ldots
\nonumber \\
&=& \int_0^\infty \prod_i ds_i \left[ 1 - R \left(  \left\{ s_i \right\} \right) w \vec{\lambda}^2 + \ldots \right] D \left( {\rm{\bf p}}^2, \left\{ s_i \right\} \right) \ ,
\end{eqnarray}
where $D \left( {\rm{\bf p}}^2, \left\{ s_i \right\} \right)$ is a usual integrand of the $\phi^3$--theory.

\subsection{Renormalization}
\label{renormalization}
We use dimensional regularization\cite{thooft_veltman_72} to compute the various diagrams obtained by decomposition. Appendix~\ref{app:computation} lines out these calculations in terms of examples. As the result of an $\epsilon$--expansion up to second order in $\epsilon^{-1}$ we obtain for the superficially divergent parts of the vertex functions
\begin{mathletters}
\label{gammafkts}
\begin{eqnarray}
\label{gammafkt2}
\Gamma_2 \left( {\rm{\bf p}} ,\vec{\lambda}; \tau, g, w \right) &=& \tau + {\rm{\bf p}}^2 + w \vec{\lambda}^2 - g^2 \frac{G_\epsilon}{\epsilon} \tau^{-\epsilon /2} \left\{ \left( 1 + \frac{\epsilon}{2} \right) \tau + \frac{1}{6} {\rm{\bf p}}^2 + \frac{5}{6} w \vec{\lambda}^2  \right\}
\nonumber \\
&+& g^4 \frac{G^2_\epsilon}{\epsilon^2} \tau^{-\epsilon} \left\{ \left( \frac{9}{4} + \frac{45}{16} \epsilon \right) \tau + \left( \frac{11}{36} + \frac{7}{432} \epsilon \right) {\rm{\bf p}}^2 + \left( \frac{65}{36} + \frac{169}{432} \epsilon \right) w \vec{\lambda}^2 \right\}
\end{eqnarray}
and
\begin{eqnarray}
\Gamma_3 \left( \left\{ {\rm{\bf 0}} , \vec{\lambda}  \right\} ; \tau, g, 0 \right) = g - 2 g^3 \frac{G_\epsilon}{\epsilon} \tau^{-\epsilon /2} + g^5 \frac{G^2_\epsilon}{\epsilon^2} \tau^{-\epsilon} \left( \frac{11}{2} + \frac{13}{8} \epsilon \right) \ ,
\end{eqnarray}
\end{mathletters}
where $G_\epsilon = (4\pi )^{-d/2}\Gamma (1 + \epsilon /2)$ with $\Gamma$ denoting the Gamma--function. We included the convergent term $-g^2 \tau^{1-\epsilon /2} G_\epsilon$ in Eq.~(\ref{gammafkt2}) since it is important in the calculation of the two loop contribution to the corresponding $Z$--factor. The vertex functions (\ref{gammafkts}) are apart from terms proportional $w \vec{\lambda}^2$ identical to those of the Potts--model. 

The $\epsilon$--poles are compensated by minimal subtraction. We employ the following renormalization scheme:
\begin{mathletters}
\begin{eqnarray}
\psi \to {\mathaccent"7017 \psi} = Z^{1/2} \psi \ ,&\quad\quad&
\tau \to {\mathaccent"7017 \tau} = Z^{-1} Z_{\tau} \tau \ ,
 \\
w \to {\mathaccent"7017 w} = Z^{-1} Z_w w \ , &\quad&
g \to {\mathaccent"7017 g} = Z^{-3/2} Z_u^{1/2} G_\epsilon^{-1/2} u^{1/2} \mu^{\epsilon /2} \ .
\end{eqnarray}
\end{mathletters}
In minimal subtraction the $Z$--factors have to be determined such that they solely cancel the $\epsilon$--poles. We find
\begin{mathletters}
\begin{eqnarray}
Z &=& 1 + \frac{1}{6}\frac{u}{\epsilon} - \frac{37}{432}\frac{u^2}{\epsilon} + \frac{11}{36}\frac{u^2}{\epsilon^2} + \mathcal O \mathnormal \left( u^3 \right) \ ,
 \\
Z_\tau &=& 1 + \frac{u}{\epsilon} - \frac{47}{48}\frac{u^2}{\epsilon} + \frac{9}{4}\frac{u^2}{\epsilon^2} + \mathcal O \mathnormal \left( u^3 \right) \ ,
 \\
Z_w &=& 1 + \frac{5}{6}\frac{u}{\epsilon} - \frac{319}{432}\frac{u^2}{\epsilon} + \frac{65}{36}\frac{u^2}{\epsilon^2} + \mathcal O \mathnormal \left( u^3 \right) \ ,
 \\
Z_u &=& 1 + 4\frac{u}{\epsilon} - \frac{59}{12}\frac{u^2}{\epsilon} + 11\frac{u^2}{\epsilon^2} + \mathcal O \mathnormal \left( u^3 \right) \ ,
\end{eqnarray}
\end{mathletters}
with $Z$, $Z_\tau$ and $Z_u$ being the known Potts--model $Z$--factors.

\section{Renormalization Group Equation And Scaling}
The unrenormalized theory has to be independent of the length scale $\mu^{-1}$ introduced by renormalization, i.e.\ 
\begin{eqnarray}
\label{independence}
\mu \frac{\partial}{\partial \mu} {\mathaccent"7017 G}_N \left( \left\{ {\rm{\bf p}} ,\vec{\lambda} \right\} ; {\mathaccent"7017 \tau}, {\mathaccent"7017 g}, {\mathaccent"7017 w} \right) = 0
\end{eqnarray}
for all $N$. Eq.~(\ref{independence}) translates via the Wilson--functions
\begin{mathletters}
\begin{eqnarray}
\label{wilson}
\beta \left( u \right) = \mu \frac{\partial u}{\partial \mu} \ ,
&\quad& 
\kappa \left( u \right) = \mu \frac{\partial
\ln \tau}{\partial \mu} \ ,
 \\
\zeta \left( u \right) = \mu \frac{\partial \ln w}{\partial \mu} \ ,
&\quad& 
\gamma \left( u \right) = \mu \frac{\partial }{\partial \mu} \ln Z
\end{eqnarray}
\end{mathletters}
(the bare quantities are kept fixed while taking the derivatives) into the Gell--Mann--Low renormalization group equation
\begin{eqnarray}
\left[ \mu \frac{\partial }{\partial \mu} + \beta \frac{\partial }{\partial u} + \tau \kappa \frac{\partial }{\partial \tau} + w \zeta \frac{\partial }{\partial w} + \frac{N}{2} \gamma \right] G_N \left( \left\{ {\rm{\bf x}} ,\vec{\lambda} \right\} ; \tau, u, w, \mu \right) = 0 \ .
\end{eqnarray}
The particular form of the Wilson--functions can be extracted from the renormalization scheme and the $Z$--factors. We find
\begin{mathletters}
\begin{eqnarray}
\beta \left( u \right) &=& u \left( \frac{7}{2}u - \frac{671}{72}u^2 - \epsilon \right) + \mathcal O \mathnormal \left( u^4 \right) \ ,
 \\
\kappa \left( u \right) &=& \frac{5}{6}u - \frac{193}{108}u^2 + \mathcal O \mathnormal \left( u^3 \right) \ ,
 \\
\zeta \left( u \right) &=& \frac{2}{3}u - \frac{47}{36}u^2 + \mathcal O \mathnormal \left( u^3 \right) \ ,
 \\
\gamma \left( u \right) &=& -\frac{1}{6}u + \frac{37}{216}u^2 + \mathcal O \mathnormal \left( u^3 \right) \ .
\end{eqnarray}
\end{mathletters}
The renormalization group equation is solved by the method of characteristics. The characteristics read
\begin{mathletters}
\begin{eqnarray}
l \frac{\partial \bar{\mu}}{\partial l} = \bar{\mu} \quad \mbox{with} \quad \bar{\mu}(1)=\mu \ ,
 \\
l \frac{\partial \bar{u}}{\partial l} = \beta \left( \bar{u}(l) \right) \quad \mbox{with} \quad \bar{u}(1)=u \ ,
 \\
l \frac{\partial}{\partial l} \ln \bar{\tau} = \kappa \left( \bar{u}(l) \right) \quad \mbox{with} \quad \bar{\tau}(1)=\tau \ ,
 \\
l \frac{\partial}{\partial l} \ln \bar{w} = \zeta \left( \bar{u}(l) \right) \quad \mbox{with} \quad \bar{w}(1)=w \ ,
 \\
l \frac{\partial}{\partial l} \ln \bar{Z} = \gamma \left( \bar{u}(l) \right) \quad \mbox{with} \quad \bar{Z}(1)=1 \ .
\end{eqnarray}
\end{mathletters}
Solving the first one is trivial. For the remaining characteristics fixed point solutions are determined. The fixed point condition $\beta \left( u^\ast \right) = 0$ leads to the infrared stable fixed point
\begin{eqnarray}
u^\ast = \frac{2}{7}\epsilon + \frac{671}{3087}\epsilon^2 + \mathcal O \mathnormal \left( \epsilon^3  \right) \ .
\end{eqnarray}
We obtain as fixed point solution of the renormalization group equation
\begin{eqnarray}
\label{fixed_point_sol}
G_N \left( \left\{ {\rm{\bf x}} ,\vec{\lambda} \right\} ; \tau, u, w, \mu \right) = 
l^{\gamma^\ast N/2} G_N \left( \left\{ {\rm{\bf x}} ,\vec{\lambda} \right\} ; \tau l^{\kappa^\ast}, u^\ast, w l^{\zeta^\ast}, \mu l \right) \ ,
\end{eqnarray}
where $\gamma^\ast = \gamma \left( u^\ast \right)$, $\kappa^\ast = \kappa \left( u^\ast \right)$ and $\zeta^\ast = \zeta \left( u^\ast \right)$.

To get a scaling relation for the vertex functions a dimensional analysis remains to be performed. It yields
\begin{eqnarray}
\label{dim_an}
G_N \left( \left\{ {\rm{\bf x}} ,\vec{\lambda} \right\} ; \tau, u, w, \mu \right) = 
\mu^{(d-2)N/2} G_N \left( \left\{ \mu {\rm{\bf x}} ,\mu^{-1}w^{1/2}\vec{\lambda} \right\} ; \mu^{-2}\tau , u, 1, 1 \right) \ .
\end{eqnarray}
From Eq.~(\ref{fixed_point_sol}) and Eq.~(\ref{dim_an}) we derive the scaling relation
\begin{eqnarray}
\label{scaling}
G_N \left( \left\{ {\rm{\bf x}} ,\vec{\lambda} \right\} ; \tau, u, w, \mu \right) = 
l^{(d-2+\eta)N/2} G_N \left( \left\{ l{\rm{\bf x}} ,l^{\zeta^\ast/2-1} \vec{\lambda} \right\} ; l^{-1/\nu}\tau , u^\ast, w, \mu \right) 
\end{eqnarray}
with the well known critical exponents for percolation
\begin{eqnarray}
\eta = \gamma^\ast = -\frac{1}{21}\epsilon - \frac{206}{9261}\epsilon^2 + \mathcal O \mathnormal \left( \epsilon^3  \right)
\end{eqnarray}
and
\begin{eqnarray}
\nu = \left( 2 - \kappa^\ast \right)^{-1} = \frac{1}{2} + \frac{5}{84}\epsilon + \frac{589}{37044}\epsilon^2 + \mathcal O \mathnormal \left( \epsilon^3  \right) \ .
\end{eqnarray}

$\phi$ remains to be determined. A glance at Eq.~(\ref{reshuffled}) shows that
\begin{eqnarray}
M^1_R \sim \vec{\lambda}^{-2} \ .
\end{eqnarray}
The scaling properties of $\vec{\lambda}$ can be deduced from Eq.~(\ref{scaling}) and hence
\begin{eqnarray}
M^1_R \sim |{\rm{\bf x}} - {\rm{\bf x}}^\prime |^{2-\zeta^\ast} \ .
\end{eqnarray}
Thus we finally obtain
\begin{eqnarray}
\phi = \nu \left( 2 - \zeta^\ast \right) = 1 + \frac{1}{42}\epsilon + \frac{4}{3087}\epsilon^2 + \mathcal O \mathnormal \left( \epsilon^3  \right) 
\end{eqnarray}
in conformity with the result by Harris and Lubensky.

\section{Rational Approximation}
\label{pade}
Since the exact value of $\phi$ is known to be unity not only in $d \geq 6$ dimensions but also in one dimension it suggests itself to do a rational approximation. The feature $\phi (d=1) = 1$ is incorporated by rewriting $\phi$ as
\begin{eqnarray}
\label{padeeq}
\phi = 1 + \epsilon \left( \frac{1}{42} + \frac{4}{3087}\epsilon - \frac{187}{154350} \epsilon^2 + \mathcal O \mathnormal \left( \epsilon^3  \right) \right)
\end{eqnarray}
being identical up to second order to the $\epsilon$--expansion result. In Fig.~\ref{plotPhi} we compare the analytic result to numerical estimates for $d=2$ by Grassberger\cite{grassberger_98} and $d=3$ by Gingold and Lobb\cite{gingold_lobb_90}. The extrapolated value $\phi (d=2)=1.04$ deviates from the simulation result $0.9825 \pm 0.008$ by roughly $5\%$. In $d=3$ one obtains $\phi (d=3)=1.05$ compared to the numerical estimate $1.117\pm 0.019$. Here the deviation is about $6\%$. 

The good agreement should be taken with caution. Due to the rich structure of the exponent $\eta$ in the percolation problem, the exponent $\psi = \gamma_w^\ast = \phi / \nu - 2 + \eta$ might be better suited for comparison to simulations. Rational approximation of $\psi$ yields
\begin{eqnarray}
\psi =  \epsilon \left( -\frac{5}{21} - \frac{187}{3087}\epsilon + \frac{334}{15435}\epsilon^2 + \mathcal O \mathnormal \left( \epsilon^3  \right) \right) \ .
\end{eqnarray}
This is compared to the simulations in Fig.~\ref{plotPsi}. The agreement for $d=3$ is reasonable. As expected the discrepancy increases for $d$ decreasing. For $d=2$ the analytic result looks unrealistic. The structure of $\psi$ appears to be too rich to be reproduced at low dimensions by a series of a few terms.

\section{Conclusions And Outlook}
\label{conclusions}
We have presented a study of randomly diluted resistor networks based on an approach by Stephen. The motivation has been twofold. Firstly we wanted to verify a result for the resistance crossover exponent $\phi$ obtained by Lubensky and Wang using a different approach. Our result for $\phi$ is in absolute agreement to that by Lubensky and Wang. Secondly we wanted to simplify the theory of random resistor networks. We demonstrated how a decomposition scheme leeds to a new interpretation of the involved Feynman diagrams: they may themselves be viewed as resistor networks. The new interpretation of the Feynman diagrams greatly improves the handling of the calculations. 

We are positive that our formulation will foster future investigations in transport on percolation clusters. As we will report shortly in a separate publication, it proved to be advantageous for nonlinear random resistor networks for which $V \sim I^r$. In the limit $r \to +0$ the resistance between two points becomes essentially equal to the length of the shortest paths between these points. Our result for the exponent for this so--called chemical distance $d_{\mbox{{\scriptsize min}}} = 2 - \epsilon /6 - \left[ 937/588 + 45/49 \left( \ln 2 -9/10 \ln 3 \right)\right] \left( \epsilon /6 \right)^2 + \mathcal O \mathnormal \left( \epsilon^3  \right)$ verifies a previous calculation by Janssen\cite{janssen_85}. The limit $r \to \infty$ is related to the red (singly connected) bonds. Our two loop calculation gives unity for the corresponding exponent in accordance with results by Blumenfeld and Aharony\cite{blumenfeld_aharony_85} and de Arcangelis {\sl et al}.\cite{arcangelis_85}. Moreover our formulation enabled us to calculate the percolation backbone exponent $D_B$ to third order in $\epsilon$ by considering the limit $r \to -1$. We find $D_B = 2 + \epsilon /21 - 172 \epsilon^2 /9261 + 2 \left( - 74639 + 22680 \zeta \left( 3 \right) \right)\epsilon^3 /4084101$, where $\zeta$ denotes the Riemann zeta function.

\acknowledgments
We acknowledge support by the Sonderforschungsbereich 237 ``Unordnung und gro{\ss}e Fluktuationen'' of the Deutsche Forschungsgemeinschaft.

\appendix
\section{Evaluation Of Diagrams}
\label{app:computation}
In this Appendix we want to sketch the computation of the Feynman diagrams. The evaluation of contributions to $\Gamma_3$ is straightforward. By virtue of $\delta =0$ one can set external momenta and $w \vec{\lambda}$'s equal to zero. Hence all diagrams resulting from one bold diagram are giving the same contribution and the decomposition reduces to a mere factor. An example is given in Fig.~\ref{dreibein}. The diagrams obtained in this fashion are the usual ones found in the literature on the Potts--model\cite{amit_76} and can be evaluated by standard procedures\cite{amit_84}.

Now we turn to $\Gamma_2$. Since $\Gamma_2$ is superficially divergent with degree $\delta =2$ a Taylor expansion up to first order in ${\rm{\bf p}}^2$ and $w \vec{\lambda}^2$ is sufficient. Contributions of order zero and proportional to ${\rm{\bf p}}^2$ are again standard. Thus we restrict ourself to demonstrate the computation of contributions to $\Gamma_2$ proportional to $w \vec{\lambda}^2$.

We revisit our example of section~\ref{2loop} and start with Eq.~(\ref{decomp_3}). As we have concluded the sum over $\vec{\kappa}$ merely gives a factor unity. After a completion of squares in the momenta the momentum integrations are straightforward. We get
\begin{eqnarray}
\label{A1}
\mbox{(\ref{decomp_3})} = \frac{g^4}{2} \frac{1}{(4\pi)^d} \int_0^\infty \prod_{i=1}^5 ds_i \frac{\exp \left( -\tau \sum_{i=1}^5 s_i \right) }{\left[ \left( s_4 + s_5 \right) \left( s_1 + s_2 + s_3 + s_4 \right) - s_4^2 \right]^{d/2} } \exp \left( - s_1 w \vec{\lambda}^2 \right)  \ .
\end{eqnarray}
At this stage it is useful to perform a change of variables: $s_4 \to t_1$, $s_5 \to t_2$, $s_1 \to t_3 x$, $s_2 \to t_3 y$ and $s_3 \to t_3 (1-x-y)$. In these variables the integral reads
\begin{eqnarray}
\label{A2}
\mbox{(\ref{A1})} = \frac{g^4}{2} \frac{1}{(4\pi)^d} \int_0^\infty dt_1 dt_2 dt_3 \int_0^1 dx \int_0^{1-x} dy \ t_3^2 \frac{\exp \left( -\tau \left( t_1 + t_2 + t_3 \right) \right) }{\left[ t_3 t_1 + t_3 t_2 + t_1 t_2 \right]^{d/2} }
 \exp \left( - t_3 x w \vec{\lambda}^2 \right)  \ .
\end{eqnarray}
After expansion for small $w \vec{\lambda}^2$ the integrations with respect to $x$ and $y$ are easily carried out. We omit the term of order zero and obtain
\begin{eqnarray}
\label{A3}
\mbox{(\ref{A2})} = - \frac{g^4}{2} \frac{w \vec{\lambda}^2}{(4\pi)^d} \int_0^\infty dt_1 dt_2 dt_3 \ \frac{1}{6} \frac{\exp \left( -\tau \left( t_1 + t_2 + t_3 \right) \right) }{\left[ t_3 t_1 + t_3 t_2 + t_1 t_2 \right]^{d/2} } t_3^3  \ .
\end{eqnarray}
Eq.~(\ref{A3}) can be expressed in terms of a parameter integral 
\begin{eqnarray}
\label{M1}
M^1 \left( a, b, c \right) &=&
\int_{{\rm{\bf p}} ,{\rm{\bf q}}} \frac{1}{ \left( a + {\rm{\bf p}}^2 \right) \left( b + {\rm{\bf q}}^2 \right) \left( c + \left( {\rm{\bf p}} + {\rm{\bf q}} \right)^2 \right) }
\nonumber \\
&=& \frac{1}{(4\pi)^d} \int_0^\infty dt_1 dt_2 dt_3  
\frac{\exp \left( - \left( at_1 + bt_2 + ct_3 \right) \right) }{\left[ t_3 t_1 + t_3 t_2 + t_1 t_2 \right]^{d/2} } \ .
\end{eqnarray}
by taking partial derivatives with respect to $a$, $b$ or $c$ at $a=b=c=\tau$. This parameter integral was introduced by Breuer and Janssen\cite{breuer_janssen_81}. They find in dimensional regularization
\begin{eqnarray}
\label{Mb}
M^1 \left( a, b, c \right) &=& \frac{G_\epsilon^2}{6\epsilon} \left\{ \left( \frac{1}{\epsilon} + \frac{25}{12} \right) \left( a^{3-\epsilon} + b^{3-\epsilon} + c^{3-\epsilon} \right) \right.
\nonumber \\
&-& \left. \left( \frac{3}{\epsilon} + \frac{21}{4} \right) \left[ a^{2-\epsilon} \left( b + c \right) + b^{2-\epsilon} \left( a + c \right) + c^{2-\epsilon} \left( a + b \right) \right] - 3abc \right\} \ .
\end{eqnarray}
In terms of this parameter integral Eq.~(\ref{A3}) reads
\begin{eqnarray}
\label{A4}
\mbox{(\ref{A3})} = - w \vec{\lambda}^2 \frac{g^4}{2} \left\{ - \frac{1}{6}  \left. \frac{\partial^3 M^1}{\partial c^3} \right|_{a=b=c=\tau} \right\} \ .
\end{eqnarray}
and is easily evaluated yielding
\begin{eqnarray}
\label{A5}
\mbox{(\ref{A4})} = - w \vec{\lambda}^2 \frac{g^4}{2} \frac{G_\epsilon^2}{\epsilon} \tau^{-\epsilon} \left\{ - \frac{1}{6\epsilon} - \frac{3}{8} \right\}  \ .
\end{eqnarray}

\begin{figure}[h]
\centerline{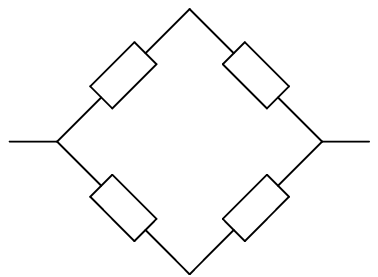}
\caption[]{\label{resistorNet}}
\end{figure}

As a second example we take the rightmost diagram in the first line of Fig.~\ref{decompRound1_2}. This diagram corresponds to the resistor network in Fig.~\ref{resistorNet}. The total resistance of this network is
\begin{eqnarray}
R \left( s_1, s_2, s_3, s_4 \right) = \frac{\left( s_1 + s_2 \right) \left( s_3 + s_4 \right) }{\left( s_1 + s_2 + s_3 + s_4 \right) }
\end{eqnarray}
Hence the saddle point evaluation of 
\begin{eqnarray}
\label{example2_1}
& &\frac{g^4}{2} \sum_{\vec{\kappa}} \int_{{\rm{\bf k}} {\rm{\bf q}}} \int_0^\infty \prod_{i=1}^5 ds_i \exp \left( -\tau \sum_{i=1}^5 s_i - \left( s_1 + s_3 \right) {\rm{\bf k}}^2 - s_5 {\rm{\bf q}}^2 \right)
\nonumber \\ 
&\times& \exp \left( - \left( s_2 + s_4 \right) \left( {\rm{\bf k}} + {\rm{\bf q}} \right)^2 - \left( s_1 + s_2 \right) w \vec{\kappa}^2 - \left( s_3 + s_4 \right) w \left( \vec{\kappa} + \vec{\lambda} \right)^2  
\right) 
\end{eqnarray}
gives
\begin{eqnarray}
\label{example2_2}
\mbox{(\ref{example2_1})} &=& - \frac{g^4}{2} \frac{w \vec{\lambda}^2}{(4\pi)^d} \int_0^\infty \prod_{i=1}^5 ds_i \frac{\left( s_1 + s_2 \right) \left( s_3 + s_4 \right) }{\left( s_1 + s_2 + s_3 + s_4 \right)}
\nonumber \\
&\times& \frac{\exp \left( -\tau \sum_{i=1}^5 s_i \right) }{\left[ s_5 \left( s_1 + s_2 + s_3 + s_4 \right) + \left( s_2 + s_4 \right) \left( s_1 + s_3 \right) \right]^{d/2} } \ ,
\end{eqnarray}
where we have already carried out the momentum integrations and the expansion for small $w \vec{\lambda}^2$. The change of variables $s_2 \to t_1 x$, $s_4 \to t_1 (1-x)$, $s_1 \to t_2 y$, $s_3 \to t_2 (1-y)$ and $s_5 \to t_3$ recasts the integral into
\begin{eqnarray}
\label{example2_3}
\mbox{(\ref{example2_2})} &=& - \frac{g^4}{2} \frac{ w \vec{\lambda}^2}{(4\pi)^d} \int_0^\infty dt_1 dt_2 dt_3 \int_0^1 dx dy \ t_1 t_2 \frac{\exp \left( -\tau \left( t_1 + t_2 + t_3 \right) \right) }{\left[ t_3 t_1 + t_3 t_2 + t_1 t_2 \right]^{d/2} }
\nonumber \\
&\times& \frac{\left( t_1 x + t_2 y \right) \left( t_1 \left( 1-x \right) + t_2 \left( 1-y \right) \right) }{t_1 + t_2} \ .
\end{eqnarray}
Carrying out the integrations with respect to $x$ and $y$ yields 
\begin{eqnarray}
\label{example2_4}
\mbox{(\ref{example2_3})} = - \frac{g^4}{2} \frac{w \vec{\lambda}^2}{(4\pi)^d} \int_0^\infty dt_1 dt_2 dt_3 \ \frac{t_1 t_2}{t_1 + t_2} \frac{\exp \left( -\tau \left( t_1 + t_2 + t_3 \right) \right) }{\left[ t_3 t_1 + t_3 t_2 + t_1 t_2 \right]^{d/2} } \left\{ \frac{1}{6}t_1^2 + \frac{1}{6}t_2^2 + \frac{1}{2}t_1 t_2 \right\} \ .
\end{eqnarray}
Since a similar structure emerges in several diagrams we introduce an additional parameter integral
\begin{eqnarray}
\label{Ma}
M^2 \left( a, b, c \right) = \frac{1}{(4\pi)^d} \int_0^\infty dt_1 dt_2 dt_3  
\frac{t_1 t_2}{t_1 + t_2} \frac{\exp \left( - \left( at_1 + bt_2 + ct_3 \right) \right) }{\left[ t_3 t_1 + t_3 t_2 + t_1 t_2 \right]^{d/2} } \ .
\end{eqnarray}
We calculate $ M^2 \left( a, b, c \right) $ in dimensional regularization which yields
\begin{eqnarray}
\label{M2}
M^2 \left( a, b, c \right) = \frac{G_\epsilon^2}{2\epsilon} \left\{ \left( \frac{1}{\epsilon} + \frac{5}{4} \right) \left( a^{2-\epsilon} + b^{2-\epsilon}\right) + \frac{1}{3} \left( \frac{1}{\epsilon} + \frac{19}{12} \right) c^{2-\epsilon} +  ab + \frac{1}{2} \left( a+b \right) c \right\} \ .
\end{eqnarray}
Now we can evaluate (\ref{example2_4}) by taking derivatives of $ M^2 \left( a, b, c \right) $:
\begin{eqnarray}
\label{example2_5}
\mbox{(\ref{example2_4})} = - w \vec{\lambda}^2 \frac{g^4}{2} \left\{ \frac{1}{3} \left. \frac{\partial^2 M}{\partial a^2} \right|_{a=b=c=\tau} + \frac{1}{2} \left. \frac{\partial^2 M}{\partial a \partial b} \right|_{a=b=c=\tau} \right\} \ .
\end{eqnarray}
We finally obtain
\begin{eqnarray}
\label{example2_6}
\mbox{(\ref{example2_5})} = - w \vec{\lambda}^2 \frac{g^4}{2} \frac{G_\epsilon^2}{\epsilon} \tau^{-\epsilon} \left\{ \frac{1}{3\epsilon} + \frac{1}{6} \right\} \ .
\end{eqnarray}

The remaining diagrams two loop diagrams contributing to $\Gamma_2$ can be treated in a similar fashion. However the two diagrams in Fig.~\ref{decompRound1_2} containing only light propagators are calculated most conveniently by observing that momenta and replica variables play exactly the same role in these diagrams (see Fig.~\ref{convCal}). Appendix~\ref{app:results} gives an overview of the two loop diagrams contributing to $\Gamma_2$ in terms of $M^1$ and $M^2$. 

\section{The Diagrams in Terms of Parameter Integrals}
\label{app:results}
\begin{figure}[h]
\begin{eqnarray*}
\raisebox{-7.25mm}{\input{./twoloop1.pstex_t}} &=& 2 \frac{d-6}{d} M^1_{3,2,1} + \frac{8}{d} \tau M^1_{4,2,1} + \frac{2}{d} \tau M^1_{3,3,1} + \frac{2}{d} I^2_3 
\\
\raisebox{-7.25mm}{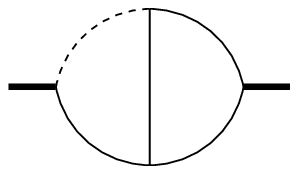} &=& \frac{4}{3} M^1_{3,2,1} + \frac{1}{3} M^2_{2,1,2} 
\\
\raisebox{-7.25mm}{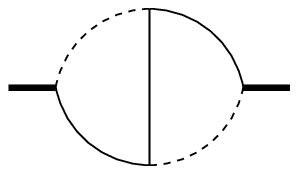} &=& 2 M^1_{3,2,1} + M^1_{2,2,2} 
\\
\raisebox{-7.25mm}{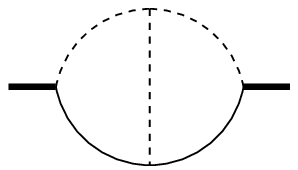} &=& 2 M^1_{3,2,1} 
\\
\raisebox{-7.25mm}{\input{./twoloop6.pstex_t}} &=& \frac{d-4}{d} M^1_{4,1,1} + \frac{4}{d} \tau M^1_{5,1,1}
\\
\raisebox{-7.25mm}{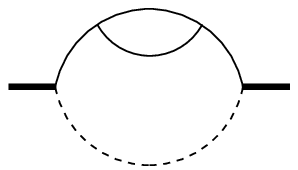} &=& 2 M^1_{4,1,1} + M^2_{1,1,3} \\
\raisebox{-7.25mm}{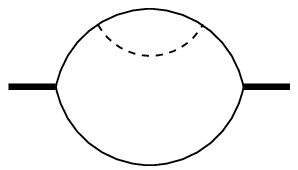} &=& \frac{1}{2} M^1_{4,1,1} + \frac{1}{6} M^2_{3,1,1}
\\
\raisebox{-7.25mm}{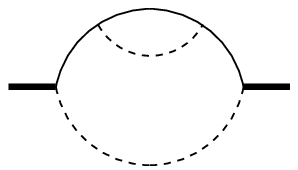} &=& 2 M^1_{4,1,1} + M^1_{3,2,1} \end{eqnarray*}
\caption[]{\label{table}}
\end{figure}

Here we list our results for the diagrams contributing to $\Gamma_2$, as far as not given in Appendix~\ref{app:computation}. For convenience we use the notation
\begin{eqnarray}
\label{notation1}
M^{1,2}_{i,j,l} = \left. \frac{(-1)^{i+j+l-3}}{(i-1)!(j-1)!(l-1)!} \frac{\partial^{i+j+l-3}}{\partial a^{i-1}\partial b^{j-1}\partial c^{l-1}}
M^{1,2} \left( a,b,c \right) \right|_{a=b=c=\tau}
\end{eqnarray}
and
\begin{eqnarray}
\label{notation2}
I_3 = \int_{{\rm{\bf p}} } \frac{1}{\left( \tau + {\rm{\bf p}}^2 \right)^3} \ .
\end{eqnarray}
Those parts of the diagrams proportional to $w \vec{\lambda}^2$ are displayed in Fig.~\ref{table}. The remaining parts can be inferred from literature on the Potts--model.


\newpage
\begin{figure}[h]
\end{figure}
\begin{figure}[h]
\begin{eqnarray*}
\raisebox{-1.3mm}{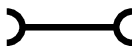}
= \hspace*{1.5mm}
\raisebox{-1.3mm}{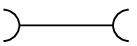}
- \hspace*{1.5mm}
\raisebox{-1.3mm}{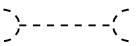} 
\end{eqnarray*}
\caption[]{\label{propagator}}
\end{figure}
\begin{figure}[h]
\begin{eqnarray*}
\Gamma_2\left( {\rm{\bf p}} ,\vec{\lambda}; \tau, g, w \right) &=& \raisebox{-7.5mm}{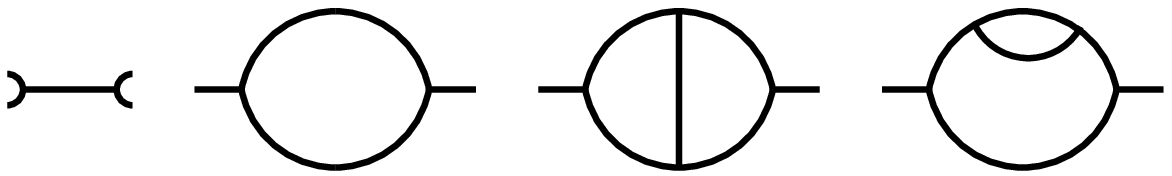}
\nonumber \\ 
\Gamma_3\left( \left\{ {\rm{\bf 0}} ,\vec{\lambda} \right\}; \tau, g, 0 \right) &=& \raisebox{-9.5mm}{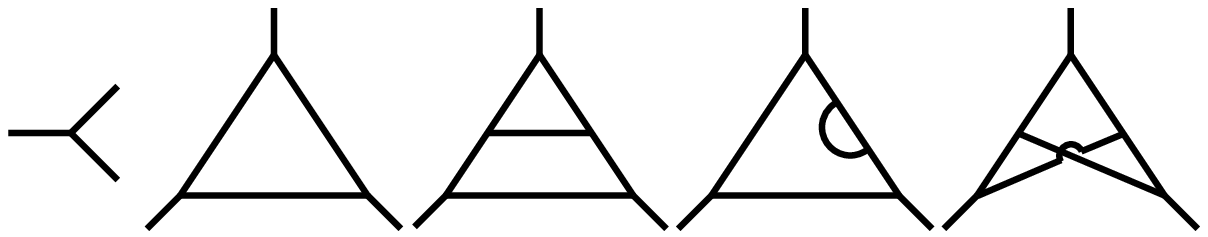} 
\end{eqnarray*}
\caption[]{\label{gammas}}
\end{figure}
\begin{figure}[h]
\begin{eqnarray*}
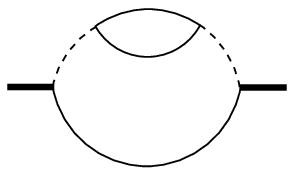
\end{eqnarray*}
\caption[]{\label{feature}}
\end{figure}
\begin{figure}[h]
\begin{eqnarray*}
\sum_{\left\{ \vec{\kappa} \right\} } \ \raisebox{-8.5mm}{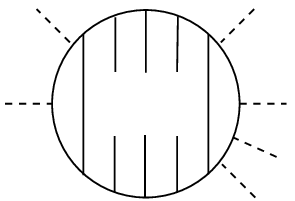} = \ \raisebox{-8.5mm}{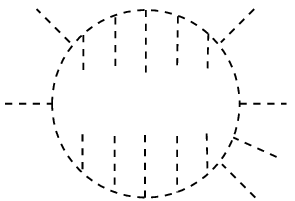}
\end{eqnarray*}
\caption[]{\label{rule}}
\end{figure}
\begin{figure}[h]
\centerline{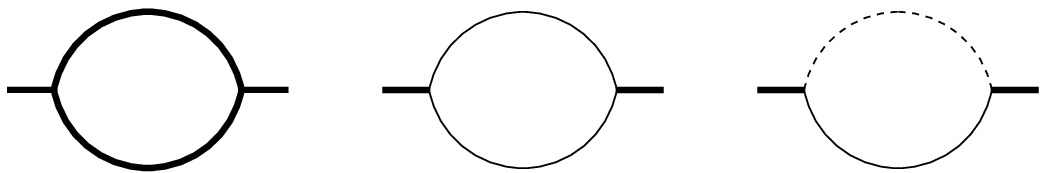}
\caption[]{\label{decompRound0}}
\end{figure}
\begin{figure}[h]
\centerline{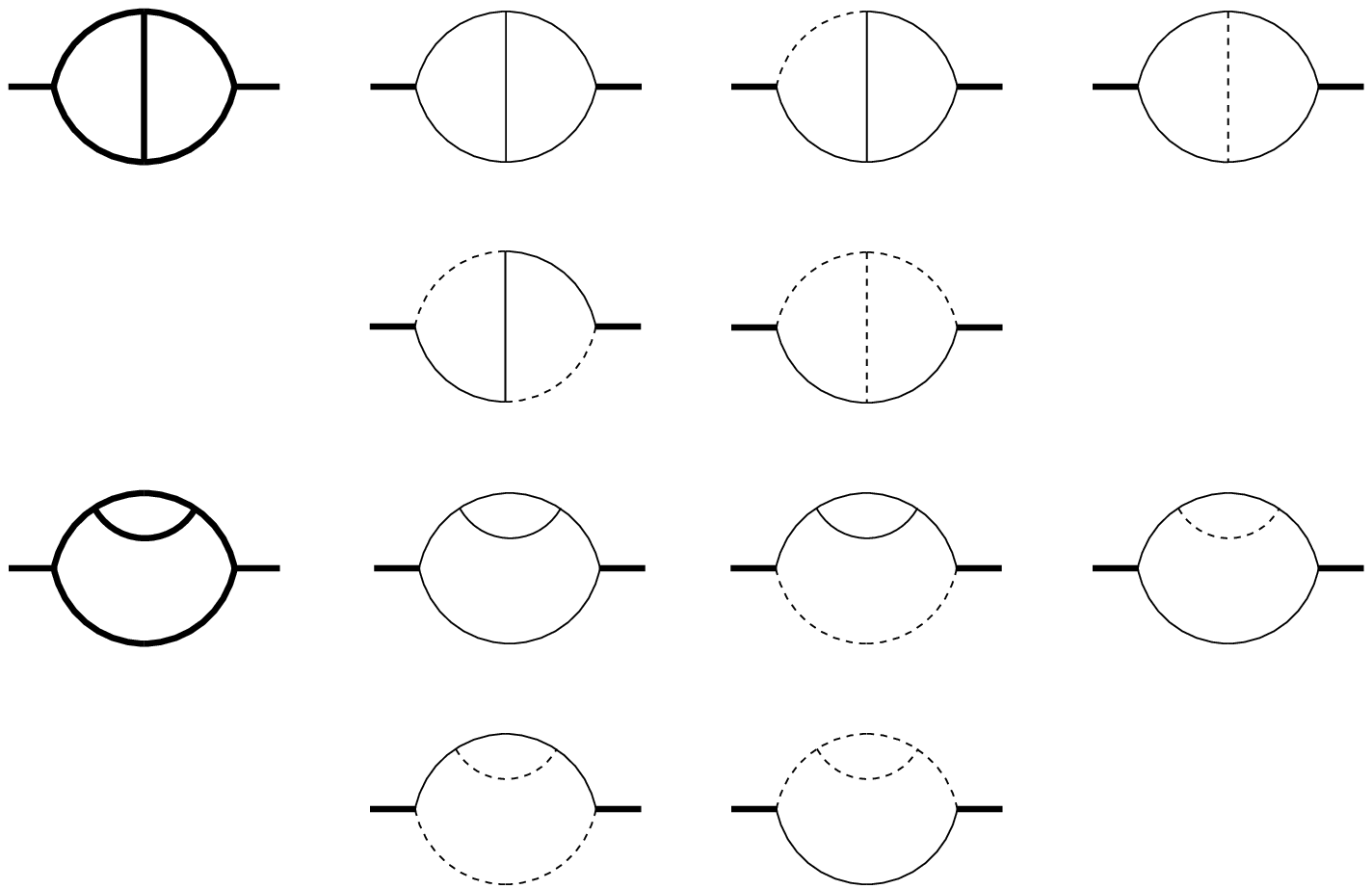}
\caption[]{\label{decompRound1_2}}
\end{figure}
\begin{figure}[h]
\centerline{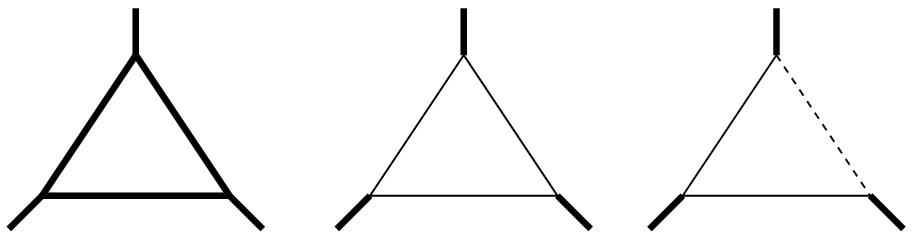}
\caption[]{\label{decompTri1}}
\end{figure}
\begin{figure}[h]
\centerline{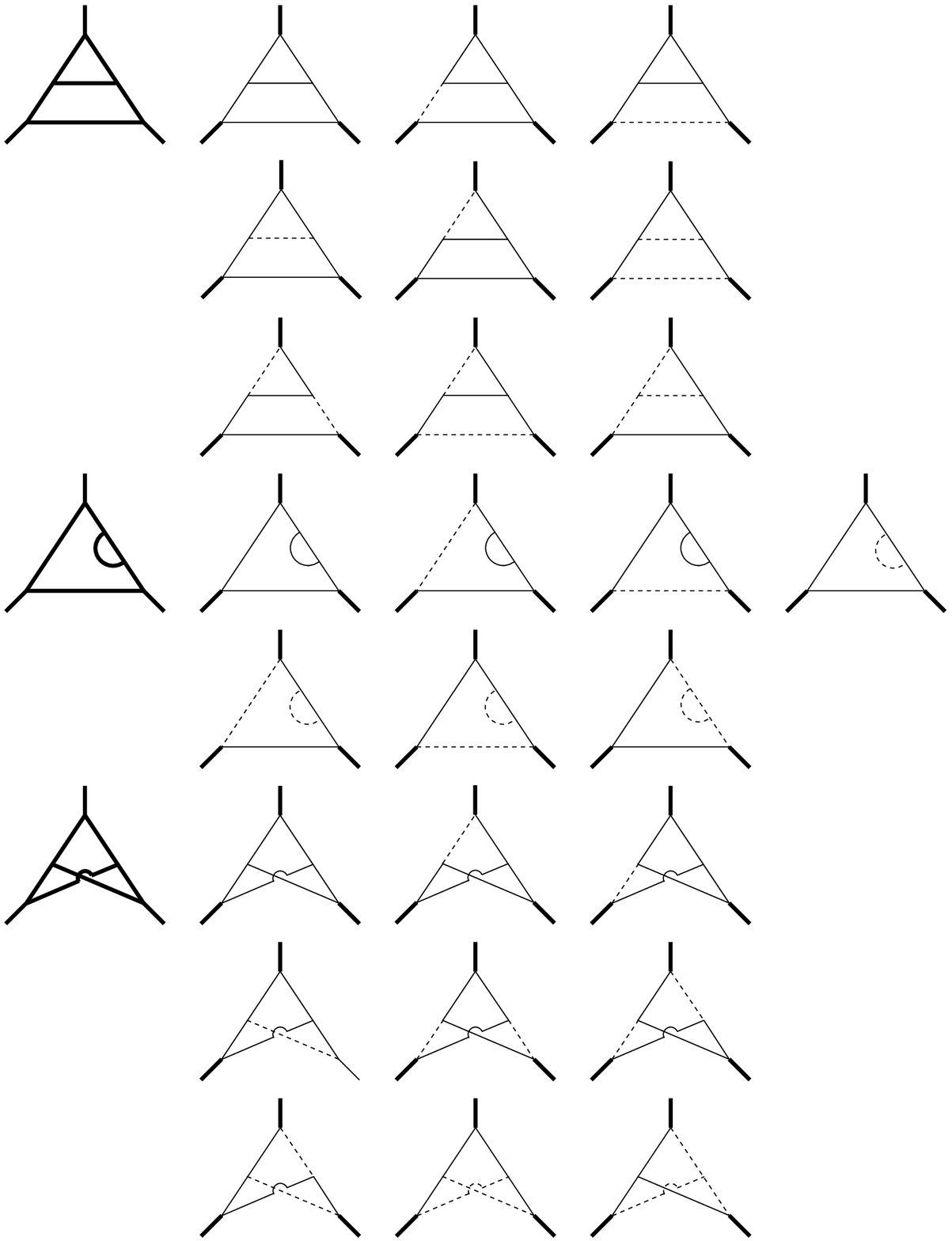}
\caption{\label{decompTri2_3_4}}
\end{figure}
\begin{figure}[h]
\epsfxsize=8.4cm
\centerline{\epsffile{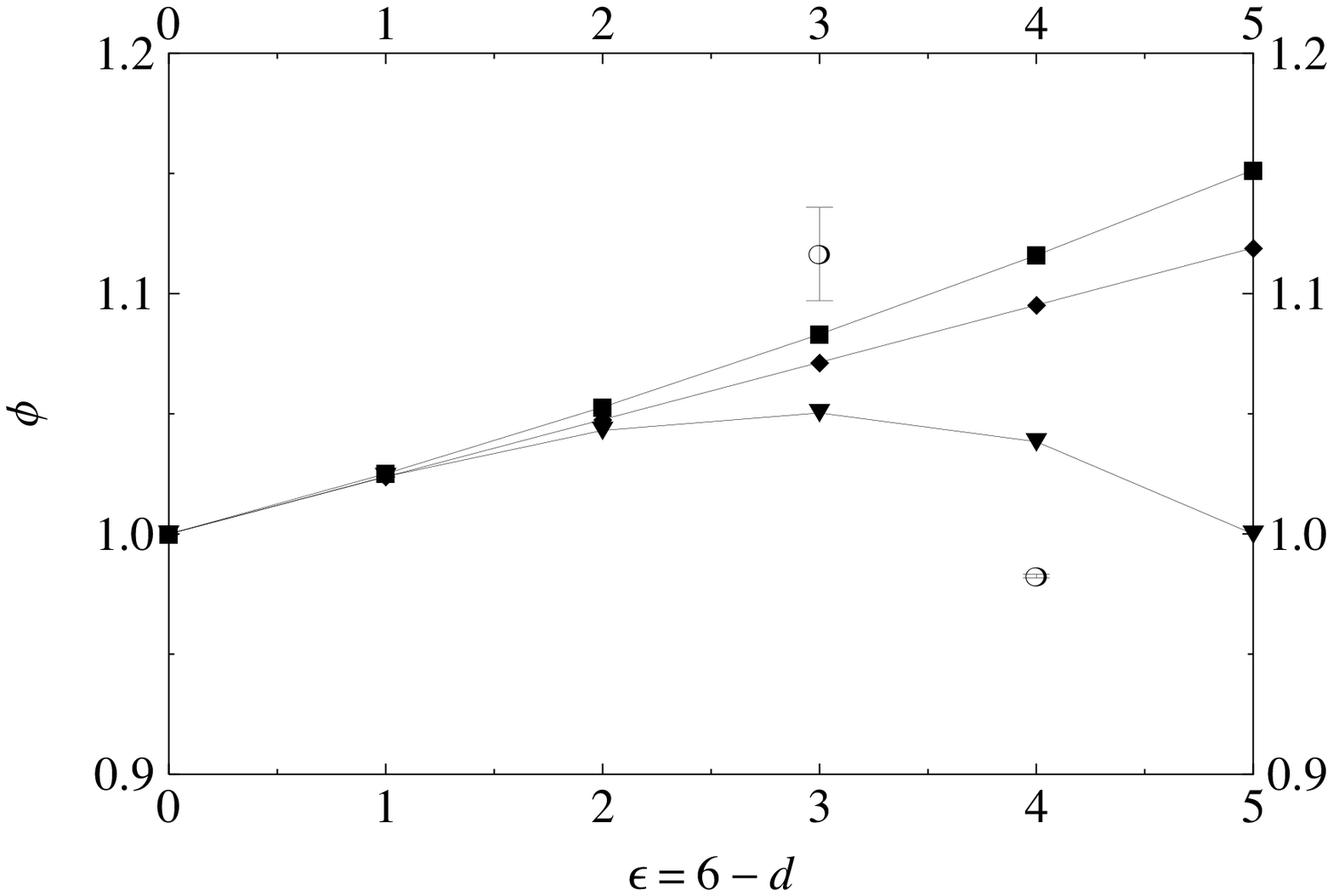}}
\caption[]{\label{plotPhi}}
\end{figure}
\begin{figure}[h]
\epsfxsize=8.4cm
\centerline{\epsffile{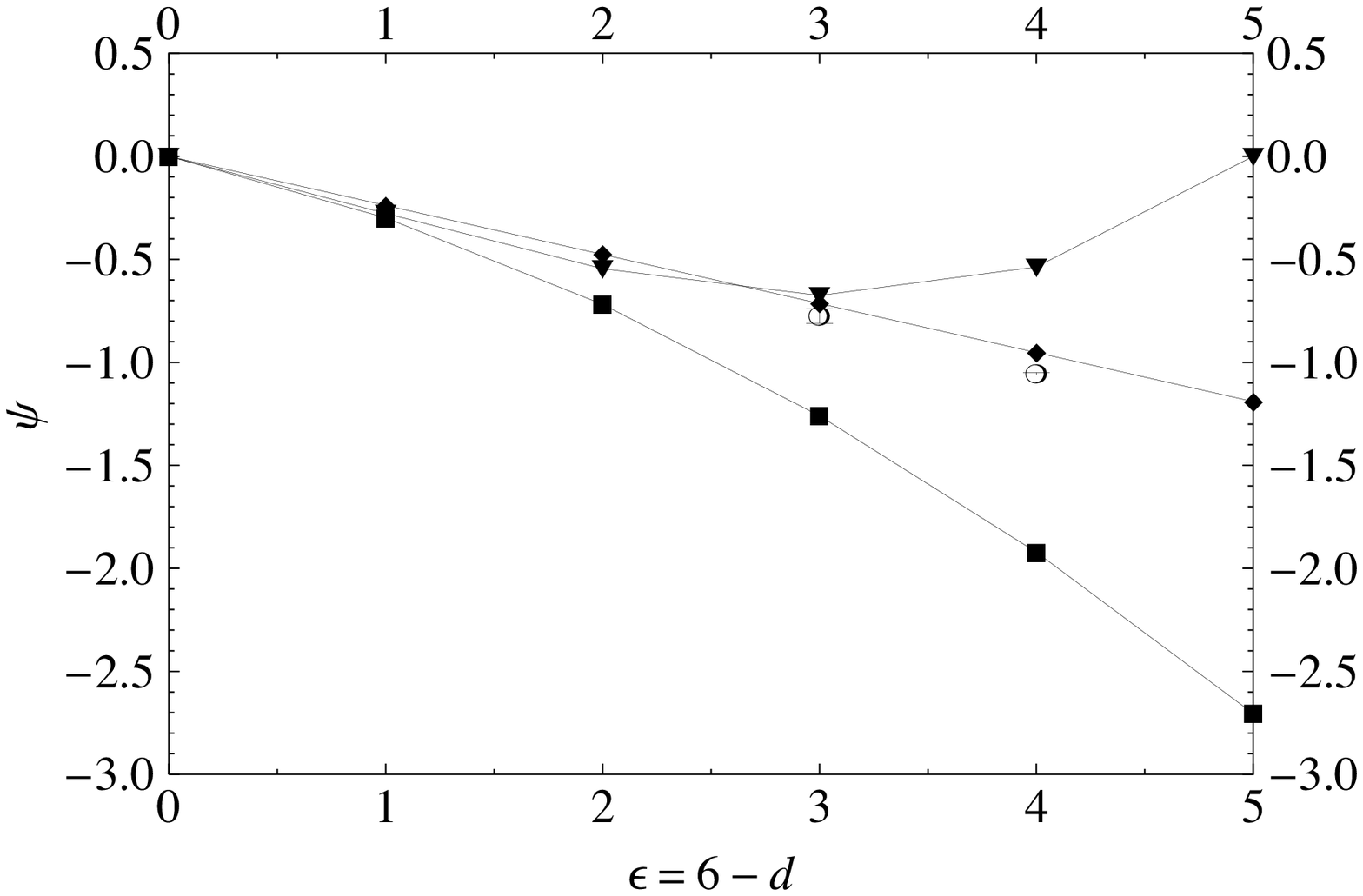}}
\caption[]{\label{plotPsi}}
\end{figure}
\begin{figure}[h]
\begin{eqnarray*}
\raisebox{-9mm}{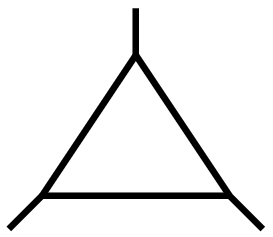} = - 2 \raisebox{-9mm}{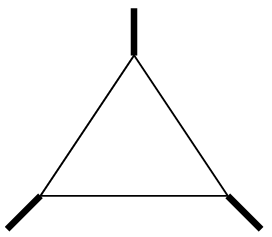} 
\end{eqnarray*}
\caption[]{\label{dreibein}}
\end{figure}
\begin{figure}[h]
\begin{eqnarray*}
\raisebox{-7.25mm}{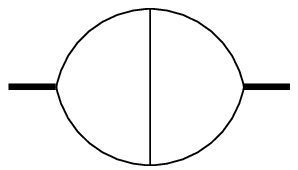} &=& w \vec{\lambda}^2 \left. \frac{\partial}{\partial {\rm{\bf p}}^2} \raisebox{-7.25mm}{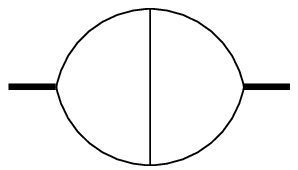} \right|_{{\rm{\bf p}} ={\rm{\bf 0}}}
\\
\raisebox{-7.25mm}{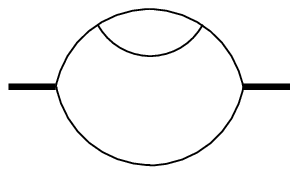} &=& w \vec{\lambda}^2 \left. \frac{\partial}{\partial {\rm{\bf p}}^2} \raisebox{-7.25mm}{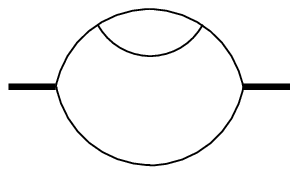} \right|_{{\rm{\bf p}} ={\rm{\bf 0}}}
\end{eqnarray*}
\caption[]{\label{convCal}}
\end{figure}

\noindent
FIG.~\ref{propagator}\newline
The principal propagator (bold) decomposes into two propagators. One of them (light) is carrying currents and we refer to it as conducting. The other one (dashed) is not carrying $\vec{\lambda}$'s and we call it insulating.
\newline\vspace{0.4cm}\\
FIG.~\ref{gammas}\newline
The superficially divergent vertex functions $\Gamma_2$ and $\Gamma_3$. Terms of $\mathcal O \mathnormal$(3 loop) have been neglected.
\newline\vspace{0.4cm}\\
FIG.~\ref{feature}\newline
A diagram encountered by decomposing the contributions to $\Gamma_2$.
\newline\vspace{0.4cm}\\
FIG.~\ref{rule}\newline
An important feature of the decomposition of the bold diagrams. By applying the decomposition scheme one may obtain sub--diagrams which are connected to the rest of their diagram solely by insulating legs. The hatched blob on the left hand side stands for such a sub--diagram. For each closed loop of conducting propagators a sum over an independent $\vec{\kappa}$ has to be performed. In the limit $D \to 0$ these sums merely produce factors unity. Thus all conducting propagators can be replaced by insulating ones and there are no summations necessary in the sub--diagram. 
\newline\vspace{0.4cm}\\
FIG.~\ref{decompRound0}\newline
Decomposition of the one loop diagram contributing to $\Gamma_2$.
\newline\vspace{0.4cm}\\
FIG.~\ref{decompRound1_2}\newline
Decomposition of the two loop diagrams contributing to $\Gamma_2$.
\newline\vspace{0.4cm}\\
FIG.~\ref{decompTri1}\newline
Decomposition of the one loop diagram contributing to $\Gamma_3$.
\newline\vspace{0.4cm}\\
FIG.~\ref{decompTri2_3_4}\newline
Decomposition of the two loop diagrams contributing to $\Gamma_3$.
\newline\vspace{0.4cm}\\
FIG.~\ref{plotPhi}\newline
Dependence of the exponent $\phi$ on dimensionality. The $\epsilon$--expansion  up first (diamonds) and second order (squares) as well as the rational approximation (triangles) are compared to numerical results (circles) by Grassberger and Gingold and Lobb. They determined the exponent $t$ for the conductivity $\Sigma$, $\Sigma \sim | p-p_c |^t$, by simulations. $t$ is related to $\phi$ via $\phi = t - (d-2)\nu$. In $d=2$ $\nu$ is known exactly\cite{nijs_79,nienhuis_82}: $\nu = 4/3$. For $d=3$ we use a Monte Carlo result by Ziff and Stell\cite{ziff_stell}:  $\nu = 0.875 \pm 0.008$.
\newline\vspace{0.4cm}\\
FIG.~\ref{plotPsi}\newline
Dependence of the exponent $\psi= (t - \gamma ) /\nu + 2 -d$ on dimensionality. As in Fig.~\ref{plotPhi} diamonds and squares refer to the $\epsilon$--expansion, triangles to the rational approximation and circles to numerical results by Grassberger and Gingold and Lobb. For the exponent $\gamma$ governing the mass of finite clusters we use $\gamma = 43/18$\cite{nijs_79,nienhuis_82} for $d=2$ and $\gamma = 1.795 \pm 0.005$\cite{ziff_stell} for $d=3$.
\newline\vspace{0.4cm}\\
FIG.~\ref{dreibein}\newline
Since $\Gamma_3$ is superficially divergent with $\delta =0$ it is sufficient to evaluate the diagrams at zero external momenta and currents. Thus all diagrams resulting from one bold three leg diagram are giving the same contribution. The decomposition results in a mere factor, being identical to the tensor contraction in the Potts--model.
\newline\vspace{0.4cm}\\
FIG.~\ref{dreibein}\newline
The resistor network corresponding to the rightmost diagram in the first line of Fig.~\ref{decompRound1_2}. The conducting propagators are interpreted as conductors whereas the insulating propagators are interpreted as open bonds. The Schwinger parameters $s_i$ correspond to resistances.
\newline\vspace{0.4cm}\\
FIG.~\ref{convCal}\newline
A convenient way to extract the contributions proportional to $w \vec{\lambda}^2$ from diagrams not containing insulating propagators.
\newline\vspace{0.4cm}\\
FIG.~\ref{table}\newline
Listing of contributions to the diagrammatic expansion proportional to $w \vec{\lambda}^2$. The right hand sides remain to be multiplied by a factor $- w \vec{\lambda}^2 G^2_\epsilon /\epsilon$ which we dropped for notational simplicity.
%
\newpage


\begin{references}
\bibitem{bunde_havlin_91_stauffer_aharony_92} For a review see e.g.\ A. Bunde and S. Havlin, {\em Fractals and Disordered Systems} (Springer, Berlin, 1991); D. Stauffer and A. Aharony, {\em Introduction to Percolation Theory} (Taylor{\&}Francis, London, 1992).
\bibitem{harris_fisch_77}A. B. Harris and R. Fisch, Phys.\ Rev.\ Lett.\ {\bf 38}, 796 (1977).
\bibitem{dasgupta_harris_lubensky_78}C. Dasgupta, A. B. Harris and T. C. Lubensky, Phys.\ Rev.\ B {\bf 17}, 1375 (1978).
\bibitem{rammal_lemieux_tremlay_85}R. Rammal, M.-A. Lemieux and A.-M. S. Tremblay, Phys.\ Rev.\ Lett.\ {\bf54}, 1087 (1985).
\bibitem{harris_lubensky_85}A. B. Harris and T. C. Lubensky, Phys.\ Rev.\ Lett.\ {\bf 54}, 1088 (1985).
\bibitem{last_thouless_71}B. J. Last and D. J. Thouless, Phys.\ Rev.\ Lett.\ {\bf 27}, 1719 (1971).
\bibitem{skal_shklovskii_74}A. S. Skal and B. I. Shklovskii, Fiz.\ Tekh.\ Poluprovodn.\ {\bf 8}, 1582 (1974) [Sov.\ Phys.--Semicond.\ {\bf 8}, 1029 (1975)].
\bibitem{straley_76}J. P. Straley, J.\ Phys.\ C {\bf 9}, 783 (1976).
\bibitem{ephross_shklovskii_76}A. L. Ephross and B. I. Shklovskii, Phys.\ Status Solidi {\bf 76}, 475 (1976).
\bibitem{degennes_76}P. G. de Gennes, J.\ Phys.\ Lett.\ (Paris) {\bf 37}, L1 (1976).
\bibitem{straley_77}J. P. Straley, Phys.\ Rev.\ B {\bf 15}, 5733 (1977).
\bibitem{alexander_orbach_82}S. Alexander and R. Orbach, J.\ Phys.\ Lett.\ (Paris) {\bf 43}, L625 (1982).
\bibitem{fortuin_kasteleyn_72}C. M. Fortuin and P. W. Kasteleyn, Physica (Utrecht) {\bf 57}, 536 (1972); P. W. Kasteleyn and C. M. Fortuin, J.\ Phys.\ Soc.\ Jpn.\ Suppl.\ {\bf 26}, 11 (1969).
\bibitem{stephen_78}M. J. Stephen, Phys.\ Rev.\ B {\bf 33}, 4444 (1978).
\bibitem{harris_lubensky_87a}A. B. Harris and T. C. Lubensky, Phys.\ Rev.\ B {\bf 35}, 6987 (1987).
\bibitem{lubensky_wang_85}T. C. Lubensky and J. Wang, Phys.\ Rev.\ B {\bf 33}, 4998 (1985).
\bibitem{harris_lubensky_87b}A. B. Harris and T. C. Lubensky, Phys.\ Rev.\ B {\bf 35}, 6964 (1987).
\bibitem{john_lubensky_85}S. John and T. C. Lubensky, Phys.\ Rev.\ Lett.\ {\bf 55}, 1014 (1985); Phys.\ Rev.\ B {\bf 34}, 4815 (1986).
\bibitem{lubensky_tremblay_86}T. C. Lubensky and A.-M. S. Tremblay, Phys.\ Rev.\ B {\bf 34}, 3408 (1986).
\bibitem{park_harris_lubensky_87}Y. Park, A. B. Harris and T. C. Lubensky, Phys.\ Rev.\ B {\bf 35}, 5048 (1987).
\bibitem{harris_87}A. B. Harris, Phys.\ Rev.\ B {\bf 35}, 5056 (1987).
\bibitem{fourcade_tremblay_95}B. Fourcade and A.-M. S. Tremblay, Phys.\ Rev.\ E {\bf 51}, 4095 (1995).
\bibitem{mezard_parizi_virasoro_87}See e.g.\ M. Mezard, G. Parizi and M. A. Virasoro {\em Spin Glass Theory and Beyond} (World Scientific, Singapore, 1987).
\bibitem{Zia_Wallace_75} See e.g.\ R. K. P. Zia and D. J. Wallace, J.\ Phys.\ A {\bf 8}, 1495 (1975).
\bibitem{thooft_veltman_72} G. t'Hooft and H. Veltman, Nucl.\ Phys.\ B {\bf 44}, 189 (1972).
\bibitem{janssen_85} H. K. Janssen, Z.\ Phys.\ B {\bf 58}, 311 (1985).
\bibitem{blumenfeld_aharony_85} R. Blumenfeld and A. Aharony, J.\ Phys.\ A {\bf 18}, L443 (1985). 
\bibitem{arcangelis_85} L. de Arcangelis, S. Redner and A. Coniglio, J.\ Phys.\ A {\bf 18}, L805 (1985).
\bibitem{amit_76} See e.g.\ D. J. Amit, J.\ Phys.\ A {\bf 9}, 1441 (1976).
\bibitem{amit_84} See e.g.\ D. J. Amit {\em Field Theory, the Renormalization Group, and Critical Phenomena} (World Scientific, Singapore, 1984); J. Zinn--Justin {\em Quantum Field Theory and Critical Phenomena} (Clarendon, Oxford, 1989).
\bibitem{breuer_janssen_81} N. Breuer and H. K. Janssen, Z.\ Phys.\ B {\bf 41}, 55 (1981).
\bibitem{grassberger_98} P. Grassberger, cond--mat/9808095.
\bibitem{gingold_lobb_90} D. B. Gingold and C. J. Lobb, Phys.\ Rev.\ B {\bf 42}, 8220 (1990).
\bibitem{nijs_79} M. P. M. den Nijs, J.\ Phys.\ A {\bf 12}, 1857 (1979).
\bibitem{nienhuis_82} B. Nienhuis, J.\ Phys.\ A {\bf 15}, 199 (1982).
\bibitem{ziff_stell} R. M. Ziff and G. Stell, unpublished; P. N. Strenski, R. M. Bradley and J.--M. Debierre, Phys.\ Rev.\ Lett.\ {\bf 66}, 1330 (1991).
\end{references}
\end{document}